\documentclass[aps,eqsecnum,preprint,floats,epsf,epsfig,nofootinbib,letter]{revtex4}
\usepackage{amsfonts}
\usepackage{amsmath}
\usepackage{amssymb}
\usepackage{mathptmx}
\usepackage[dvips]{color}
\usepackage{epsfig}
\usepackage{graphicx}
\begin{document}
\def\be{\begin{eqnarray}}
\def\en{\end{eqnarray}}
\def\non{\nonumber}
\def\la{\langle}
\def\ra{\rangle}
\def\nc{N_c^{\rm eff}}
\def\vp{\varepsilon}
\def\drho{\bar\rho}
\def\deta{\bar\eta}
\def\A{{\cal A}}
\def\B{{\cal B}}
\def\c{{\cal C}}
\def\d{{\cal D}}
\def\e{{\cal E}}
\def\p{{\cal P}}
\def\t{{\cal T}}
\def\CP{{\it CP}~}
\def\up{\uparrow}
\def\dw{\downarrow}
\def\vma{{_{V-A}}}
\def\vpa{{_{V+A}}}
\def\smp{{_{S-P}}}
\def\spp{{_{S+P}}}
\def\J{{J/\psi}}
\def\ov{\overline}
\def\Lqcd{{\Lambda_{\rm QCD}}}
\def\pr{{Phys. Rev.}~}
\def\prl{{Phys. Rev. Lett.}~}
\def\pl{{Phys. Lett.}~}
\def\np{{Nucl. Phys.}~}
\def\zp{{Z. Phys.}~}
\def\lsim{ {\ \lower-1.2pt\vbox{\hbox{\rlap{$<$}\lower5pt\vbox{\hbox{$\sim$}
}}}\ } }
\def\gsim{ {\ \lower-1.2pt\vbox{\hbox{\rlap{$>$}\lower5pt\vbox{\hbox{$\sim$}
}}}\ } }
\newcommand{\acp}{\ensuremath{A_{CP}}}


\font\el=cmbx10 scaled \magstep2{\obeylines\hfill October, 2009}

\vskip 1.5 cm

\centerline{\large\bf QCD Factorization for Charmless Hadronic $B_s$ Decays Revisited}
\bigskip
\centerline{\bf Hai-Yang Cheng,$^{1,2}$ Chun-Khiang Chua$^3$}
\medskip
\centerline{$^1$ Institute of Physics, Academia Sinica}
\centerline{Taipei, Taiwan 115, Republic of China}
\medskip
\medskip
\centerline{$^2$ Physics Department, Brookhaven National
Laboratory} \centerline{Upton, New York 11973}
\medskip
\medskip
\centerline{$^3$ Department of Physics, Chung Yuan Christian University}
\centerline{Chung-Li, Taiwan 320, Republic of China}
\medskip

\bigskip
\bigskip
\centerline{\bf Abstract}
\bigskip
\small

Branching fractions and {\it CP}-violating asymmetries of charmless $\bar B_s\to PP,~VP,~VV$ decays ($P$ and $V$ denoting pseudoscalar and vector mesons, respectively) are re-examined in the framework of QCD factorization (QCDF). We take into account subleading power corrections to the penguin annihilation topology and to color-suppressed tree amplitudes that are crucial for resolving the \CP puzzles and rate deficit problems with penguin-dominated two-body decays  and color-suppressed tree-dominated $\pi^0\pi^0$ and $\rho^0\pi^0$ modes
in the $B_{u,d}$ sector. Many of the $B_s\to h_1h_2$ decays can be related to $B_d\to h_1h_2$ ones via $U$-spin or SU(3) symmetry. Some useful model-independent relations can be derived and tested.
Mixing-induced \CP asymmetries for many of the penguin-dominated decays are predicted to be very small in the standard model. They are sensitive to New Physics and offer rich possibilities of new discoveries.
Measurements of direct {\it CP}-violating asymmetries can be used to discriminate QCDF from other competing approaches such as pQCD and soft-collinear effective theory.

\eject
\section{Introduction}
The phenomenology of nonleptonic two-body decays of $B$ mesons offers rich opportunities for our understanding of the underlying mechanism for hadronic weak decays and \CP violation. In the past decade, nearly 100 charmless decays of $B_{u,d}$ mesons  have been observed at $B$ factories, BaBar and Belle, with
a statistical significance of at least four standard deviations (for a review, see \cite{ChengSmith}). The CDF Collaboration has made unique contributions to the measurements of charmless hadronic $B_s$ decays. Recently, Belle has also started to study the weak decays of the $B_s$ meson.

Many of the $B_s\to h_1h_2$ decays can be related to $B_d\to h'_1h'_2$ ones via $U$-spin or SU(3) symmetry. Some useful model-independent relations can be derived and tested. For example, direct \CP asymmetries of $\bar B_s\to K^+\pi^-$ and $\bar B_d\to K^-\pi^+$ are related to each other by $U$-spin symmetry. Therefore, the use of flavor symmetry will be  helpful to control the hadronic uncertainties in  $\bar B_s\to h_1h_2$ decay amplitudes.

Analogous to the neutral $B_d$ system, \CP violation in $B_s$ decays also occurs through the interference of decay amplitudes with and without $B_s-\overline B_s$ mixing. It is known that the mixing-induced \CP violation of $B_d\to J/\psi K$ is governed by $\sin 2\beta$. Likewise, the decay $B_s\to J/\psi\phi$ is the benchmark in the $B_s$ system with mixing-induced \CP asymmetry characterized by $\sin 2\beta_s$. In the standard model (SM), the phase $\beta_s$ is very small, of order 1 degree. Consequently, $B_s\to J/\psi\phi$ and several charmless penguin-dominated $B_s$ decays e.g. $\bar B_s\to K^{(*)0}\bar K^{(*)0},\eta^{(')}\eta^{(')},\phi\phi$ are the ideal places to search for New Physics as \CP violation from physics beyond the SM can compete or even dominate over the small SM \CP phase. Recently, both CDF \cite{CDFbetas} and D0 \cite{D0betas} have observed 1-2 $\sigma$ deviations from the SM prediction for $\beta_s$. Because of the possibilities of new discoveries, the search for New Physics in the $B_s$ system will be the main focus of the forthcoming experiments at Fermilab, LHCb and Super $B$ factories.

Theoretically, two-body $B_s$ decays have been studied in the framework of generalized factorization  \cite{ChengBs}, QCD factorization (QCDF) \cite{QCDFBs,BN,DescotesGenon,BRY,Bartsch}, perturbative QCD (pQCD) \cite{pQCDBs,AliBs,Xiao} and soft-collinear effective theory (SCET) \cite{Zupan,SCETVP}. In this work we will re-examine and update the QCDF predictions. Especially, we shall pay attention to the issue of power corrections. From the study of charmless hadronic $B_{u,d}$ decays, we learned that two subleading $1/m_b$ power corrections are needed in QCDF in order to account for the observed rates and \CP asymmetries. Power corrections to the penguin annihilation topology, corresponding to the so-called ``scenario S4" in \cite{BN}, are crucial for accommodating the branching fractions of penguin-dominated $B_{u,d}\to PP,VP,VV$ decays on the one hand and direct \CP asymmetries of $\bar B_d\to K^-\pi^+$, $\bar B_d\to K^{*-}\pi^+$, $B^-\to K^-\rho^0$ and $\bar B_d\to\pi^+\pi^-$ on the other hand. Otherwise, the predicted rates will be too small and {\it CP}-violating asymmetries of above-mentioned modes will be wrong in signs when confronted with experiment. However, power corrections due to penguin annihilation will bring new \CP puzzles for the decays $B^-\to K^-\pi^0,~K^-\eta,\pi^-\eta$, $\bar B_d\to \bar K^{*0}\eta$ and $\bar B_d\to\pi^0\pi^0$: Signs of their $\acp$'s are flipped into the wrong ones when compared with experiment. It has been shown in \cite{CCcp} that soft corrections to the color-suppressed tree amplitude due to spectator scattering and/or final-state interactions will bring the  aforementioned \CP asymmetries to the right track and accommodate the observed $\pi^0\pi^0$ and $\rho^0\pi^0$ rates simultaneously. \footnote{It is well known that a large complex electroweak penguin can also solve the $B\to K\pi$ \CP puzzle with the difference of $\acp(B^-\to \bar K^0\pi^-)$ and $\acp(\bar B^0\to K^-\pi^+)$ (see e.g. \cite{CSKim}). Since the electroweak penguin amplitude $P_{\rm EW}$ is essentially real in the standard model, one needs New Physics to produce new {\it strong} and {\it weak} phases for $P_{\rm EW}$. In principle, it will be difficult to discriminate between large complex color-suppressed tree $C$ and large $P_{\rm EW}$ scenarios in the penguin-dominated decays. However, as pointed out in \cite{CCBud}, the two schemes can lead to very distinct predictions for tree-dominated decays where $P_{EW}\ll C$. The observed decay rates of $\bar B^0\to \pi^0\pi^0,\rho^0\pi^0$ and the \CP puzzles with $\pi^-\eta$ and $\pi^0\pi^0$ can be
resolved by a large complex $C$ but not $P_{\rm EW}$. In the $B_{u,d}$ sector, there are 13 modes in which \CP asymmetries have been measured with significance above $1.8\sigma$: $K^-\pi^+,\pi^+\pi^-,K^-\eta,\bar K^{*0}\eta,K^-\rho^0, \rho^\pm\pi^\mp$ and $\rho^+K^-,K^{*-}\pi^+,K^-\pi^0, \pi^-\eta,\omega\bar K^0,\pi^0\pi^0,\rho^-\pi^+$. We have shown in \cite{CCBud} that the QCDF predictions of $\acp$ for aforementioned 13 decays are in agreement with experiment except the decay $\bar B^0\to \omega \bar K^0$. However, we notice that BaBar and Belle measurements of  $\acp(\omega\bar K^0)$ are opposite in sign.
}
Recently we have given a detailed study of charmless hadronic $B_{u,d}\to PP,VP,VV$ decays within the framework of QCDF incorporating aforementioned power corrections \cite{CCBud}.
In this work we shall generalize the study to $B_s$ decays. So far $\bar B_s\to K^+\pi^-$ is the only hadronic decay mode in the $B_s$ sector that its direct \CP violation has been measured \cite{CDFcp}. The resulting \CP asymmetry $\acp(\bar B_s\to K^+\pi^-)=0.39\pm0.17$ differs from zero by $2.2\sigma$ deviations. Just as the decay $\bar B_d\to K^-\pi^+$, the predicted \CP asymmetry for $\bar B_s\to K^+\pi^-$ in the heavy quark limit is wrong in sign and too small in magnitude. As we shall see below, we need penguin annihilation to get the right sign and magnitude for $\acp(\bar B_s\to K^+\pi^-)$.

This work is organized as follows. We outline the QCDF framework in Sec. 2 and specify various input parameters, such as form factors, light-cone distribution amplitudes and the parameters for power corrections in Sec. 3. Then $B_s\to PP,VP,VV$ decays are analyzed in details in Secs. 4, 5 and 6, respectively.
Conclusions are given in Sec. 7.

\section{$B$ decays in QCD factorization}
Within the framework of QCD factorization \cite{BBNS}, the effective
Hamiltonian matrix elements are written in the form
\begin{equation}\label{fac}
   \langle M_1M_2 |{\cal H}_{\rm eff}|\overline B\rangle
  \! =\! \frac{G_F}{\sqrt2}\sum_{p=u,c} \! \lambda_p^{(q)}\,
\!   \langle M_1M_2 |{\cal T_A}^{h,p}\!+\!{\cal
T_B}^{h,p}|\overline B\rangle \,,
\end{equation}
where $\lambda_p^{(q)}\equiv V_{pb}V_{pq}^*$ with $q=d,s$, and the
superscript $h $ denotes the helicity of the final-state meson.
For $PP$ and $VP$ final states, $h=0$. ${\cal
T_A}^{h,p}$ describes contributions from naive factorization, vertex
corrections, penguin contractions and spectator scattering expressed
in terms of the flavor operators $a_i^{p,h}$, while ${\cal T_B}$
contains annihilation topology amplitudes characterized by  the
annihilation operators $b_i^{p,h}$. Specifically \cite{BBNS}
\be
{\cal T_A}^h &=& a_1^p(M_1M_2)\delta_{pu}(\bar u b)_\vma\otimes (\bar qu)_\vma + a_2^p(M_1M_2)\delta_{pu}(\bar q b)_\vma\otimes (\bar uu)_\vma   \non \\
&+& a_3^p(M_1M_2)\sum(\bar q b)_\vma\otimes (\bar q'q')_\vma + a_4^p(M_1M_2)\sum(\bar q' b)_\vma\otimes (\bar qq')_\vma   \non \\
&+& a_5^p(M_1M_2)\sum(\bar q b)_\vma\otimes (\bar q'q')_\vpa + a_6^p(M_1M_2)\sum(-2)(\bar q' b)_\smp\otimes (\bar qq')_\spp   \\
&+& a_7^p(M_1M_2)\sum(\bar q b)_\vma\otimes {3\over 2}e_q(\bar q'q')_\vpa + a_8^p(M_1M_2)\sum(-2)(\bar q' b)_\smp\otimes {3\over 2}(\bar qq')_\spp   \non \\
&+& a_9^p(M_1M_2)\sum(\bar q b)_\vma\otimes {3\over 2}e_q(\bar q'q')_\vma + a_{10}^p(M_1M_2)\sum(\bar q' b)_\vma\otimes {3\over 2}e_q(\bar qq')_\vma,  \non
\en
where $(\bar q_1q_2)_{_{V\pm A}}\equiv \bar q_1\gamma_\mu(1\pm\gamma_5)q_2$ and $(\bar q_1q_2)_{_{S\pm P}}\equiv\bar q_1(1\pm\gamma_5)q_2$ and the summation is over $q'=u,d,s$. The symbol $\otimes$ indicates that the matrix elements of the operators in ${\cal T_A}$ are to be evaluated in the factorized form. For the decays $\bar B\to PP,VP,VV$, the relevant factorizable matrix elements are
\be
\label{eq:X}
X^{(\bar B P_1,P_2)} &\equiv& \la P_2|J_{\mu}|0\ra\la P_1|J'^{\mu}|\ov B\ra=if_{P_2}(m_{B}^2-m^2_{P_1}) F_0^{ B P_1}(m_{P_2}^2),  \non \\
X^{(\bar BP,V)} &\equiv & \la V| J_{\mu}|0\ra\la
P|J'^{\mu}|\ov B \ra=2f_V\,m_Bp_c F_1^{ B
P}(m_{V}^2),   \non \\
X^{( \bar BV,P)} &\equiv &
\la P | J_{\mu}|0\ra\la V|J'^{\mu}|\ov B
\ra=2f_P\,m_Bp_cA_0^{B V}(m_{P}^2),  \non \\
X_h^{( \bar BV_1,V_2)} &\equiv & \la V_2 |J_{\mu}|0\ra\la
V_1|J'^{\mu}|\ov B \ra =- if_{V_2}m_2\Bigg[
(\vp^*_1\cdot\vp^*_2) (m_{B}+m_{V_1})A_1^{ BV_1}(m_{V_2}^2)  \non \\
&-& (\vp^*_1\cdot p_{_{B}})(\vp^*_2 \cdot p_{_{B}}){2A_2^{
BV_1}(m_{V_2}^2)\over (m_{B}+m_{V_1}) } +
i\epsilon_{\mu\nu\alpha\beta}\vp^{*\mu}_2\vp^{*\nu}_1p^\alpha_{_{B}}
p^\beta_1\,{2V^{ BV_1}(m_{V_2}^2)\over (m_{B}+m_{V_1}) }\Bigg],
\en
where we have followed the conventional definition for form factors \cite{BSW}. For $B\to VP,PV$ amplitudes, we have applied the
replacement $m_V\vp^*\cdot p_B\to
m_Bp_c$ with $p_c$ being the c.m. momentum. The longitudinal ($h=0$) and transverse ($h=\pm$) components of $X^{( \bar BV_1,V_2)}_h$ are given by
 \be \label{eq:Xh}
 X_0^{(\ov BV_1,V_2)} &=& {if_{V_2}\over 2m_{V_1}}\left[
 (m_B^2-m_{V_1}^2-m_{V_2}^2)(m_B+m_{V_1})A_1^{BV_1}(q^2)-{4m_B^2p_c^2\over
 m_B+m_{V_1}}A_2^{BV_1}(q^2)\right], \non \\
 X_\pm^{(\ov BV_1,V_2)} &=& -if_{V_2}m_Bm_{V_2}\left[
 \left(1+{m_{V_1}\over m_B}\right)A_1^{BV_1}(q^2)\mp{2p_c\over
 m_B+m_{V_1}}V^{BV_1}(q^2)\right].
 \en

The flavor operators $a_i^{p,h}$ are basically the Wilson coefficients
in conjunction with short-distance nonfactorizable corrections such
as vertex corrections and hard spectator interactions. In general,
they have the expressions \cite{BBNS,BN}
 \be \label{eq:ai}
  a_i^{p,h}(M_1M_2) &=&
 \left(c_i+{c_{i\pm1}\over N_c}\right)N_i^h(M_2)
   + {c_{i\pm1}\over N_c}\,{C_F\alpha_s\over
 4\pi}\Big[V_i^h(M_2)+{4\pi^2\over N_c}H_i^h(M_1M_2)\Big]+P_i^{h,p}(M_2),
 \en
where $i=1,\cdots,10$,  the upper (lower) signs apply when $i$ is
odd (even), $c_i$ are the Wilson coefficients,
$C_F=(N_c^2-1)/(2N_c)$ with $N_c=3$, $M_2$ is the emitted meson
and $M_1$ shares the same spectator quark with the $B$ meson. The
quantities $V_i^h(M_2)$ account for vertex corrections,
$H_i^h(M_1M_2)$ for hard spectator interactions with a hard gluon
exchange between the emitted meson and the spectator quark of the
$B$ meson and $P_i(M_2)$ for penguin contractions.   The expression
of the quantities $N_i^h(M_2)$ reads
 \be
 N_i^h(M_2)=\begin{cases} 0, & $i=6,8$, \cr
                 1, & {\rm else}. \cr \end{cases}
 \en

 The weak annihilation contributions to the decay  $\overline B\to
M_{1}M_2$ can be described in terms of the building blocks $b_i^{p,h}$ and $b_{i,{\rm EW}}^{p,h}$
\begin{eqnarray}\label{eq:h1ksann}
\frac{G_F}{\sqrt2} \sum_{p=u,c} \! \lambda_p^{(q)}\, \!\langle M_{1}M_2
|{\cal T_B}^{h,p} |\overline B^0\rangle &=&
i\frac{G_F}{\sqrt{2}}\sum_{p=u,c} \lambda_p^{(q)}
 f_B f_{M_1} f_{M_{2}}\sum_i (d_ib_i^{p,h}+d'_ib_{i,{\rm EW}}^{p,h}).
\end{eqnarray}
The building blocks have the expressions \cite{BN}
 \be \label{eq:bi}
 b_1 &=& {C_F\over N_c^2}c_1A_1^i, \qquad\quad b_3={C_F\over
 N_c^2}\left[c_3A_1^i+c_5(A_3^i+A_3^f)+N_cc_6A_3^f\right], \non \\
 b_2 &=& {C_F\over N_c^2}c_2A_1^i, \qquad\quad b_4={C_F\over
 N_c^2}\left[c_4A_1^i+c_6A_2^f\right], \non \\
 b_{\rm 3,EW} &=& {C_F\over
 N_c^2}\left[c_9A_1^{i}+c_7(A_3^{i}+A_3^{f})+N_cc_8A_3^{i}\right],
 \non \\
 b_{\rm 4,EW} &=& {C_F\over
 N_c^2}\left[c_{10}A_1^{i}+c_8A_2^{i}\right].
 \en
Here for simplicity we have omitted the superscripts $p$ and $h$ in above expressions.  The subscripts 1,2,3 of $A_n^{i,f}$ denote the annihilation
amplitudes induced from $(V-A)(V-A)$, $(V-A)(V+A)$ and $(S-P)(S+P)$ operators,
respectively, and the superscripts $i$ and $f$ refer to gluon emission from the
initial and final-state quarks, respectively. Following \cite{BN} we choose the
convention that $M_1$  contains an antiquark from the weak vertex and $M_2$
contains a quark from the weak vertex.

For the explicit expressions of vertex, hard spectator corrections and annihilation contributions, the reader is referred to \cite{BBNS,BN,BRY} for details.  The decay amplitudes of $\bar B_s\to PP,VP$ are given in Appendix A of \cite{BN} and can be easily generalized to $\bar B_s\to VV$ (see
\cite{Bartsch} for explicit expressions of $\bar B_s\to VV$ amplitudes). In practice, it is more convenient to express the decay amplitudes in terms of the flavor operators $\alpha_i^{h,p}$ and the annihilation operators  $\beta_i^p$ which are related to the coefficients $a_i^{h,p}$ and $b_i^p$ by
\begin{eqnarray}\label{eq:alphai}
   \alpha_1^{h}(M_1 M_2) &=& a_1^{h}(M_1 M_2) \,, \nonumber\\
   \alpha_2^{h}(M_1 M_2) &=& a_2^{h}(M_1 M_2) \,, \nonumber\\
   \alpha_3^{h,p}(M_1 M_2) &=& \left\{
    \begin{array}{cl}
     a_3^{h,p}(M_1 M_2) - a_5^{h,p}(M_1 M_2)
      & \quad \mbox{for~} M_1 M_2=PP, \, VP , \\
     a_3^{h,p}(M_1 M_2) + a_5^{h,p}(M_1 M_2)
      & \quad \mbox{for~} M_1 M_2=VV,\, PV ,
    \end{array}\right. \nonumber\\
   \alpha_4^{h,p}(M_1 M_2) &=& \left\{
    \begin{array}{cl}
     a_4^{h,p}(M_1 M_2) + r_{\chi}^{M_2}\,a_6^{h,p}(M_1 M_2)
      & \quad \mbox{for~} M_1 M_2=PP, \, PV , \\
     a_4^{h,p}(M_1 M_2) - r_{\chi}^{M_2}\,a_6^{h,p}(M_1 M_2)
      & \quad \mbox{for~} M_1 M_2=VP\,, VV,
    \end{array}\right.\\
   \alpha_{3,\rm EW}^{h,p}(M_1 M_2) &=& \left\{
    \begin{array}{cl}
     a_9^{h,p}(M_1 M_2) - a_7^{h,p}(M_1 M_2)
      & \quad \mbox{for~} M_1 M_2=PP, \, VP , \\
     a_9^{h,p}(M_1 M_2) + a_7^{h,p}(M_1 M_2)
      & \quad \mbox{for~} M_1 M_2=VV,\, PV ,
    \end{array}\right. \nonumber\\
   \alpha_{4,\rm EW}^{h,p}(M_1 M_2) &=& \left\{
    \begin{array}{cl}
     a_{10}^{h,p}(M_1 M_2) + r_{\chi}^{M_2}\,a_8^{h,p}(M_1 M_2)
      & \quad \mbox{for~} M_1 M_2=PP, \, PV , \\
     a_{10}^{h,p}(M_1 M_2) - r_{\chi}^{M_2}\,a_8^{h,p}(M_1 M_2)
      & \quad \mbox{for~} M_1 M_2=VP\,, VV,
     \end{array}\right. \nonumber
\end{eqnarray}
and
\be \label{eq:beta}
 \beta_i^p (M_1 M_2) =\frac{i f_B f_{M_1}
f_{M_2}}{X^{(\overline B M_1,M_2)}}b_i^p.
\en
The order of the arguments of $\alpha_i^p (M_1 M_2)$ and $\beta_i^p(M_1
M_2)$ is consistent with the order of the arguments of
$X^{(\overline B M_1, M_2)}\equiv A_{M_1M_2}$.
The chiral factor $r_\chi$ is given by
\be \label{eq:rchi}
 r_\chi^P(\mu)={2m_P^2\over m_b(\mu)(m_2+m_1)(\mu)},  \qquad r_\chi^V(\mu) = \frac{2m_V}{m_b(\mu)}\,\frac{f_V^\perp(\mu)}{f_V} \,.
\en

\section{Input parameters}
It is clear from Eq. (\ref{eq:X}) that we need the information on decay constants and form factors in order to evaluate the factorizable matrix elements of 4-quark operators. Moreover, we also need to know the light-cone distribution amplitudes of light hadrons in order to evaluate the nonfactorizable contributions.

\subsection{Form factors}

There exist one lattice and several model calculations of form factors for $B_s\to P,V$ transitions:

\renewcommand{\theenumi}{\arabic{enumi})}
\begin{enumerate}
\item
In the pQCD approach, the relevant form factors obtained at $q^2=0$ are \cite{AliBs} (for simplicity, form factors hereafter are always referred to the ones at $q^2=0$, unless specified otherwise)
\be \label{eq:FFpQCD}
&&  F_0^{B_sK}=0.24^{+0.05+0.00}_{-0.04-0.01}, \quad~~ F_0^{B_s\eta_s}=0.30^{+0.06+0.01}_{-0.05-0.01}, \non \\
&& V^{B_sK^*}=0.21^{+0.04+0.00}_{-0.03-0.01}, \quad
A_0^{B_sK^*}=0.25^{+0.05+0.00}_{-0.05-0.01}, \quad
A_1^{B_sK^*}=0.16^{+0.03+0.00}_{-0.03-0.01}, \non \\
&& V^{B_s\phi}=0.25^{+0.05+0.00}_{-0.04-0.01}, \quad~~
A_0^{B_s\phi}=0.30^{+0.05+0.00}_{-0.05-0.01}, \quad~~
A_1^{B_s\phi}=0.19^{+0.03+0.00}_{-0.03-0.01}.
\en
\item
Form factors obtained by  QCD sum rules are
\be
F_0^{B_sK}=0.30^{+0.04}_{-0.03},
\en
for the $B_s\to K$ transition \cite{Melic} and
\be \label{eq:FFSR}
&& V^{B_sK^*}=0.311\pm0.026, \quad
A_0^{B_sK^*}=0.360\pm0.034, \quad
A_1^{B_sK^*}=0.233\pm0.022, \non \\
&& V^{B_s\phi}=0.434\pm0.035, \quad~~
A_0^{B_s\phi}=0.474\pm0.033, \quad~~
A_1^{B_s\phi}=0.311\pm0.030,
\en
for $B_s\to V$ transitions \cite{Ball}.
\item
Another light-cone sum rule calculation based on heavy quark effective theory gives \cite{WuFF}
\be
F_0^{B_sK}=0.296\pm0.018, \qquad F_0^{B_s\eta}=0.281^{+0.015}_{-0.016},
\en and
\be  \label{eq:LCHQETff}
&& V^{B_sK^*}=0.285^{+0.013}_{-0.013}, \quad
A_0^{B_sK^*}=0.222^{+0.011}_{-0.010}, \quad
A_1^{B_sK^*}=0.227^{+0.010}_{-0.012}, \non \\
&& V^{B_s\phi}=0.339^{+0.016}_{-0.017}, \quad~~
A_0^{B_s\phi}=0.269^{+0.014}_{-0.014}, \quad~~
A_1^{B_s\phi}=0.271^{+0.014}_{-0.014}.
\en
It is clear that form factors obtained by sum rules
are larger than the pQCD ones.

\item
A light cone quark model in conjunction with soft collinear effective theory was constructed in \cite{LuFF}. The predictions are
\be \label{eq:FFour}
&& F_0^{B_sK}=0.290, \quad\quad F_0^{B_s\eta_s}=0.288, \non\\
&& V^{B_sK^*}=0.323, \quad~~
A_0^{B_sK^*}=0.279, \quad
A_1^{B_sK^*}=0.228, \non \\
&& V^{B_s\phi}=0.329, \qquad~
A_0^{B_s\phi}=0.279, \quad~~
A_1^{B_s\phi}=0.232\,.
\en
\item
A straightforward application of the covariant light-front quark model of \cite{CLF} yields \cite{Verma}
\be \label{eq:CLF}
&& V^{B_sK^*}=0.23, \quad~~
A_0^{B_sK^*}=0.26, \quad
A_1^{B_sK^*}=0.19, \non \\
&& V^{B_s\phi}=0.30, \qquad~
A_0^{B_s\phi}=0.32, \quad~~
A_1^{B_s\phi}=0.26\,,
\en
all with errors estimated to be $\pm0.01$\,.

\item A recent lattice QCD calculation yields $F_0^{B_sK}=0.23\pm0.05\pm0.04$ \cite{lattice}.

\end{enumerate}
For comparison,
Beneke and Neubert \cite{BN} used
\be
F_0^{B_sK}=0.31\pm0.05, \qquad
A_0^{B_sK^*}=0.29\pm0.05, \qquad
A_0^{B_s\phi}=0.34\pm0.05,
\en
and
\be \label{eq:FFetaBN}
F_0^{B_s\to\eta^{(')}}=F_0^{BK}{f_{\eta^{(')}}^q\over f_\pi}+F_2{\sqrt{2}f_{\eta^{(')}}^q+f_{\eta^{(')}}^s\over \sqrt{3}f_\pi},
\en
while Beneke, Rohrer and Yang \cite{BRY} employed
\be \label{eq:FFBRY}
A_0^{B_sK^*}=0.33\pm0.05, \quad A_0^{B_s\phi}=0.38^{+0.10}_{-0.02}.
\en

Note that it is most convenient to express the form factors for $B\to \eta^{(')}$ transitions  in terms of the flavor states   $q\bar q\equiv (u\bar u+d\bar
d)/\sqrt{2}$, $s\bar s$ and $c\bar c$ labeled by the $\eta_q$, $\eta_s$ and $\eta_{c}^0$, respectively. Neglecting the small mixing with $\eta_c^0$, we have
\be \label{eq:FFeta}
F^{B_s\eta}=-F^{B_s\eta_s}\sin\theta, \quad
F^{B_s\eta'}=F^{B_s\eta_s}\cos\theta,
\en
where $\theta$ is the $\eta_q-\eta_s$ mixing angle defined by
\be
|\eta\ra &=& \cos\theta|\eta_q\ra-\sin\theta|\eta_s\ra, \non \\
|\eta'\ra &=& \sin\theta|\eta_q\ra+\cos\theta|\eta_s\ra,
\en
with $\theta=(39.3\pm1.0)^\circ$ in the Feldmann-Kroll-Stech mixing scheme \cite{FKS}.

From the above discussions we see that the form factor $F_0^{B_sK}$ at $q^2=0$ ranges from 0.23 to 0.31\,. In the QCDF approach, if $F_0^{B_sK}(0)=0.31$ is employed, we find that the predicted branching fractions $\B(\bar B_s\to K^+\pi^-)\approx 9.1\times 10^{-6}$ and $\B(\bar B_s\to K^+K^-)\approx 34\times 10^{-6}$ will be far above the experimental measurements of $(5.0\pm1.1)\times 10^{-6}$ \cite{CDF} and $(25.7\pm 3.6)\times 10^{-6}$ \cite{Tonelli,Belle:KK}, respectively. Hence we shall use $F_0^{B_sK}(0)=0.24$ obtained by the lattice calculation. Note that a $\chi^2$ analysis by one of us (C.K.C.) with the available data of $B_s\to PP$ also yields $F_0^{B_sK}(0)=0.240^{+0.021}_{-0.007}$ \cite{Chua}. For other form factors,  we shall use $F_0^{B_s\eta_s}(0)=0.28$ and $B_s\to V$ transition form factors given by  Eq. (\ref{eq:CLF}) with some modifications on $B_s\to K^*$ ones (see Table \ref{tab:input}) .

\subsection{Decay constants}

 Decay constants of various vector mesons defined by
 \be \label{eq:decayc}
 \la V(p,\epsilon)|\bar q_2\gamma_\mu q_1|0\ra &=& -if_V m_V\epsilon^*_\mu, \non \\
   \langle V(p,\epsilon) |\bar q_2 \sigma_{\mu\nu}q_1 |0\rangle
 &=& -f_V^\perp (\epsilon_{\mu}^* p^\nu -
\epsilon_{\nu}^* p^\mu)\,,
 \en
are listed in Table \ref{tab:input}. They are taken from \cite{BallfV}. For pseudoscalar mesons, we use $f_\pi=132$ MeV and $f_K=160$ MeV. Decay constants $f^{q}_{\eta^{(')}}$, $f_{\eta^{(')}}^{s}$ and $f_{\eta^{(')}}^c$ defined by
\be
\la 0|\bar q\gamma_\mu\gamma_5q|\eta^{(')}\ra=i{1\over\sqrt{2}}f_{\eta^{(')}}^q  q_\mu, \quad \la 0|\bar s\gamma_\mu\gamma_5s|\eta^{(')}\ra=if_{\eta^{(')}}^s q_\mu, \quad \la 0|\bar c\gamma_\mu\gamma_5c|\eta^{(')}\ra=if_{\eta^{(')}}^c q_\mu
\en
are also needed in calculations.
For the decay constants $f_{\eta^{(')}}^q$ and $f_{\eta^{(')}}^s$, we shall use the values
\be   \label{fdecay}
f_{\eta}^q=107\,{\rm
MeV}, \quad f_\eta^s=-112\,{\rm MeV}, \quad f_{\eta'}^q= 89\,{\rm
MeV}, \quad f_{\eta'}^s=137\,{\rm MeV}
\en
obtained in \cite{FKS}. As for $f_{\eta^{(')}}^c$,
a straightforward perturbative calculation gives \cite{fetac}
\be
f_{\eta^{(')}}^c=-{m_{\eta^{(')}}^2\over 12 m_c^2}\,{f_{\eta^{(')}}^q\over\sqrt{2}}.
\en

\begin{table}[t]
\caption{
Input parameters.
The values of the scale dependent quantities $f^\perp_V(\mu)$ and $a^{\bot,V}_{1,2}(\mu)$ are
given for $\mu=1\,\rm{GeV}$. The values of Gegenbauer moments are taken from \cite{Ball2007} and Wolfenstein parameters  from \cite{CKMfitter}.
} \label{tab:input}
\begin{center}
\begin{tabular}{|c||c|c|c|c|c|c|c|c|}
\hline\hline
\multicolumn{7}{|c|}{Light vector mesons} \\
\hline
$V$  & $f_V({\rm MeV})$ &
 $f^\perp_V({\rm MeV})$ & $a^V_1$ & $a^V_2$ & $a^{\bot,V}_1$ & $a^{\bot,V}_2$\\
\hline
$\rho$ & $216\pm3$ & $165\pm 9$ & 0 & $0.15\pm0.07$ &  0 & $0.14\pm0.06$ \\
$\omega$ & $187\pm5$ & $151\pm 9$ & 0 & $0.15\pm 0.07$ & 0 & $0.14\pm0.06$ \\
$\phi$ & $215\pm5$ & $186\pm 9$ & 0 & $0.18\pm 0.08$ & 0 & $0.14\pm0.07$\\
$K^*$ & $220\pm5$ & $185\pm 10$ & $0.03\pm 0.02$ & $0.11\pm 0.09$ & $0.04\pm0.03$ & $0.10\pm0.08$ \\
\hline\hline
\multicolumn{7}{|c|}{Light pseudoscalar mesons} \\
\hline
 \multicolumn{2}{|c|}{$a_1^\pi$} & \multicolumn{2}{|c|}{$a_2^\pi$} &
\multicolumn{2}{|c|}{$a_1^K$} & $a_2^K$ \\
\hline
 \multicolumn{2}{|c|}{0} & \multicolumn{2}{|c|}{$0.25\pm0.15$} & \multicolumn{2}{|c|}{$0.06\pm0.03$} & $0.25\pm0.15$ \\
\hline\hline
\multicolumn{7}{|c|}{$B$ mesons} \\
\hline
$B$ & \multicolumn{2}{|c|}{$m_B({\rm GeV}$)} & $\tau_B({\rm ps})$ &
\multicolumn{2}{|c|}{$f_B({\rm MeV})$} & $\lambda_B({\rm MeV})$ \\
\hline
$B_u$ & \multicolumn{2}{|c|}{$5.279$} & $1.638$ &
\multicolumn{2}{|c|}{$210\pm 20$} & $300\pm 100$ \\
\hline
$B_d$ & \multicolumn{2}{|c|}{$5.279$} & $1.525$ &
\multicolumn{2}{|c|}{$210\pm 20$} & $300\pm 100$ \\
\hline
$B_s$ & \multicolumn{2}{|c|}{$5.366$} & $1.472$ &
\multicolumn{2}{|c|}{$230\pm 20$} & $300\pm 100$ \\
\hline\hline
\multicolumn{7}{|c|}{Form factors at $q^2=0$} \\
\hline
\multicolumn{2}{|c|}{$F^{B_sK}_0(0)$} &
$A^{B_s K^*}_0(0)$ &
$A^{B_s K^*}_1(0)$ &
$A^{B_s K^*}_2(0)$ &
\multicolumn{2}{|c|}{$V^{B_s K^*}_0(0)$}  \\
\hline
\multicolumn{2}{|c|}{$0.24$} &
$0.30\pm0.01$ &
$0.24\pm0.01$ &
$0.22\pm0.01$ &
\multicolumn{2}{|c|}{$0.28\pm0.01$}  \\
\hline
\multicolumn{2}{|c|}{$F^{B_s \eta_s}_0(0)$} &
$A^{B_s \phi}_0(0)$ &
$A^{B_s \phi}_1(0)$ &
$A^{B_s \phi}_2(0)$ &
\multicolumn{2}{|c|}{$V^{B_s \phi}_0(0)$}  \\
\hline
\multicolumn{2}{|c|}{$0.28$} &
$0.32\pm0.01$ &
$0.26\pm0.01$ &
$0.23\pm0.01$ &
\multicolumn{2}{|c|}{$0.30\pm0.01$}  \\
\hline\hline
\multicolumn{7}{|c|}{Quark masses} \\
\hline
\multicolumn{2}{|c|}{$m_b(m_b)/{\rm GeV}$} &
$m_c(m_b)/{\rm GeV}$ & \multicolumn{2}{|c|}{$m_c^{\rm pole}/m_b^{\rm pole}$} &
\multicolumn{2}{|c|}{$m_s(2.1~{\rm GeV})/{\rm GeV}$}   \\
\hline
\multicolumn{2}{|c|}{$4.2$} &
$0.91$ & \multicolumn{2}{|c|}{$0.3$} &
\multicolumn{2}{|c|}{$0.095\pm0.020$}  \\
\hline\hline
\multicolumn{7}{|c|}{Wolfenstein parameters} \\
\hline
\multicolumn{2}{|c|}{$A$} & $\lambda$ &
$\bar\rho$ & $\bar\eta$ &
\multicolumn{2}{|c|}{$\gamma$}  \\
\hline
\multicolumn{2}{|c|}{$0.8116$} & $0.2252$ &
$0.139$ & $0.341$ &
\multicolumn{2}{|c|}{$(67.8^{+4.2}_{-3.9})^\circ$}  \\
\hline
\hline
\end{tabular}
\end{center}
\end{table}

\subsection{LCDAs}
We next specify the light-cone distribution amplitudes (LCDAs) for pseudoscalar and vector mesons. The
general expressions of twist-2 LCDAs are
 \be
 \Phi_{P}(x,\mu) &=& 6x(1-x)\left[1+\sum_{n=1}^\infty
 a_n^{P}(\mu)C_n^{3/2}(2x-1)\right], \non \\
 \Phi^{V}_{\parallel}(x,\mu) &=& 6x(1-x)\left[1+\sum_{n=1}^\infty
 a_n^{V}(\mu)C_n^{3/2}(2x-1)\right],  \non \\
 \Phi^{V}_{\perp}(x,\mu) &=& 6x(1-x)\left[1+\sum_{n=1}^\infty
 a_n^{\perp,V}(\mu)C_n^{3/2}(2x-1)\right],
 \en
and twist-3 ones
 \be
&& \Phi_p(x)=1, \qquad \Phi_\sigma(x)=6x(1-x), \non \\
&&  \Phi_v(x,\mu)=3\left[2x-1+\sum_{n=1}^\infty
 a_{n}^{\bot,V}(\mu)P_{n+1}(2x-1)\right],
 \en
where $C_n(x)$ and $P_n(x)$ are the Gegenbauer and Legendre polynomials, respectively. When three-particle amplitudes are neglected, the twist-3 $\Phi_v(x)$ can be expressed in terms of $\Phi_\perp$
\be
\Phi_v(x)=\int_0^x
\frac{\Phi_\perp(u)}{\bar u}du -\int_x^1
\frac{\Phi_\perp(u)}{u}du.
\en
The normalization of LCDAs is
 \be
 \int^1_0dx \Phi_V(x)=1, \qquad \int^1_0dx \Phi_v(x)=0.
 \en
Note that the Gegenbauer moments $a_i^{(\perp),K^*}$ displayed in Table \ref{tab:input} taken from \cite{Ball2007}
are for the mesons containing a strange quark.

The integral of the $B$
meson wave function is parameterized as \cite{BBNS}
 \begin{eqnarray}
 \int_0^1 \frac{d\rho}{1-\rho}\Phi_1^B(\rho) \equiv
 \frac{m_B}{\lambda_B}\,,
 \end{eqnarray}
where $1-\rho$ is the momentum fraction carried by the light
spectator quark in the $B$ meson. We shall use $\lambda_B=300\pm100$ MeV.

For the running quark masses we shall use \cite{PDG,Xing}
 \be \label{eq:quarkmass}
 && m_b(m_b)=4.2\,{\rm GeV}, \qquad~~~~ m_b(2.1\,{\rm GeV})=4.94\,{\rm
 GeV}, \qquad m_b(1\,{\rm GeV})=6.34\,{\rm
 GeV}, \non \\
 && m_c(m_b)=0.91\,{\rm GeV}, \qquad~~~ m_c(2.1\,{\rm GeV})=1.06\,{\rm  GeV},
 \qquad m_c(1\,{\rm GeV})=1.32\,{\rm
 GeV}, \non \\
 && m_s(2.1\,{\rm GeV})=95\,{\rm MeV}, \quad~ m_s(1\,{\rm GeV})=118\,{\rm
 MeV}, \non\\
 && m_d(2.1\,{\rm GeV})=5.0\,{\rm  MeV}, \quad~ m_u(2.1\,{\rm GeV})=2.2\,{\rm
 MeV}.
 \en
Note that the charm quark masses here are smaller than the one $m_c(m_b)=1.3\pm 0.2$ GeV adopted in \cite{BN,Bartsch} and
consistent with the high precision mass determination from lattice QCD \cite{LQCDmc}:
$m_c(3\,{\rm GeV})=0.986\pm0.010$ GeV and $m_c(m_c)=1.267\pm0.009$ GeV (see also \cite{Kuhn})
Among the quarks, the strange quark gives the major theoretical uncertainty to
the decay amplitude. Hence, we will only consider the uncertainty in the
strange quark mass given by $m_s(2.1\,{\rm GeV})=95\pm20$ MeV.
Notice that for the one-loop penguin contribution, the relevant quark mass is the pole mass rather than the current one \cite{Li:2006jb}. Since the penguin loop correction is governed by the ratio of the pole masses squared $s_i\equiv(m_i^{\rm pole}/m_b^{\rm pole})^2$ and since the pole mass is meaningful only for heavy quarks, we only need to consider the ratio of $c$ and $b$ quark pole masses given by $s_c\approx (0.3)^2$.

\subsection{Penguin annihilation}
In the QCDF approach, the hadronic $B$ decay amplitude receives contributions from tree, penguin, electroweak penguin and weak annihilation topologies. In the absence of $1/m_b$ power corrections except for the chiral enhanced penguin contributions, the leading QCDF predictions encounter three major difficulties: (i) the predicted branching fractions for penguin-dominated $B\to PP,VP,VV$ decays are systematically below the measurements, (ii) direct {\it CP}-violating asymmetries for $\bar B_d\to K^-\pi^+$, $\bar B_d\to K^{*-}\pi^+$, $B^-\to K^-\rho^0$, $\bar B_d\to \pi^+\pi^-$ and $\bar B_s\to K^+\pi^-$ have signs in disagreement with experiment, and (iii) the predicted longitudinal polarization fractions in penguin-dominated $B\to VV$ decays are usually too large and do not agree with the data. This implies the necessity of introducing $1/m_b$ power corrections. Unfortunately, there are many possible $1/m_b$ power suppressed effects and they are generally nonperturbative in nature and hence not calculable by the perturbative method.

Power corrections in QCDF always involve troublesome endpoint divergences. For
example, the annihilation amplitude has endpoint divergences even at twist-2 level and the hard spectator scattering diagram at twist-3 order is power
suppressed and posses soft and collinear divergences arising from the soft
spectator quark. Since the treatment of endpoint divergences is model dependent, subleading power corrections generally can be studied only in a
phenomenological way. We shall follow \cite{BBNS} to model the endpoint divergence $X\equiv\int^1_0 dx/\bar x$ in the annihilation and hard spectator
scattering diagrams as
 \be \label{eq:XA}
 X_A=\ln\left({m_B\over \Lambda_h}\right)(1+\rho_A e^{i\phi_A}), \qquad
 X_H=\ln\left({m_B\over \Lambda_h}\right)(1+\rho_H e^{i\phi_H}),
 \en
with $\Lambda_h$ being a typical scale of order
500 MeV, and  $\rho_{A,H}$, $\phi_{A,H}$ being the unknown real parameters.

A fit to the data of $B_{u,d}\to PP,VP,PV$ and $VV$ decays yields the values of $\rho_A$ and $\phi_A$ shown in Table \ref{tab:rhoA}. Basically, it is very similar to the so-called ``S4 scenario" presented in \cite{BN}. The fitted $\rho_A$ and $\phi_A$ for $B\to VV$ decays are taken from \cite{ChengVV}. Since the penguin annihilation effects are different for $B\to VP$ and $B\to PV$ decays,
\be
&& A_1^i\approx -A_2^i\approx 6\pi\alpha_s\left[3\left(X_A^{VP}-4+{\pi^2\over 3}\right)+r_\chi^V r_\chi^P\Big((X_A^{VP})^2-2X_A^{VP}\Big)\right], \non \\
&& A_3^i\approx  6\pi\alpha_s\left[-3r_\chi^V\left((X_A^{VP})^2-2X_A^{VP}+4-{\pi^2\over 3}\right)+r_\chi^P \left((X_A^{VP})^2-2X_A^{VP}+{\pi^2\over 3}\right)\right], \non \\
&& A_3^f\approx  6\pi\alpha_s\left[3r_\chi^V(2X_A^{VP}-1)(2-X_A^{VP})-r_\chi^P \Big(2(X_A^{VP})^2-X_A^{VP}\Big)\right],
\en
for $M_1M_2=VP$ (the definition for the parameters $r_\chi^P$ and $r_\chi^V$ can be found in Eq. (\ref{eq:rchi}) below) and
\be
&& A_1^i\approx -A_2^i\approx 6\pi\alpha_s\left[3\left(X_A^{PV}-4+{\pi^2\over 3}\right)+r_\chi^V r_\chi^P\Big((X_A^{PV})^2-2X_A^{PV}\Big)\right], \non \\
&& A_3^i\approx  6\pi\alpha_s\left[-3r_\chi^P\left((X_A^{PV})^2-2X_A^{PV}+4-{\pi^2\over 3}\right)+r_\chi^V \left((X_A^{PV})^2-2X_A^{PV}+{\pi^2\over 3}\right)\right], \non \\
&& A_3^f\approx  6\pi\alpha_s\left[-3r_\chi^P(2X_A^{PV}-1)(2-X_A^{PV})+r_\chi^V \Big(2(X_A^{PV})^2-X_A^{PV}\Big)\right],
\en
for $M_1M_2=PV$, the parameters $X_A^{VP}$ and $X_A^{PV}$ are not necessarily the same. Indeed, a fit to the $B\to VP,PV$ decays yields $\rho_A^{VP}\approx 1.07$, $\phi_A^{VP}\approx -70^\circ$ and $\rho_A^{PV}\approx 0.87$, $\phi_A^{PV}\approx -30^\circ$ (see Table \ref{tab:rhoA}).
For $B_s\to PP,VP,VV$ decays, we shall assume that their default values are similar to that in $B_{u,d}$ decays as shown in Table \ref{tab:rhoA}. For the estimate of theoretical uncertainties, we shall assign an error of $\pm0.1$ to $\rho_A$ and $\pm 20^\circ$ to $\phi_A$.

\begin{table}[h!]
 \caption{The parameters $\rho_A$ and $\phi_A$ for penguin annihilation.
 } \label{tab:rhoA}
\begin{center}
 \begin{tabular}{| l c c |l c c |} \hline
 {Modes}~~~~~~~~ & $\rho_A$  \qquad &  $\phi_A$ \quad & ~~Modes ~~~~~~~~ & $\rho_A$\qquad  & $\phi_A$ \\ \hline
 $B\to PP$ & 1.10 & $-50^\circ$~~ & ~~$B_s\to PP$ & 1.00 & $-55^\circ$ \\
 $B\to VP$ & 1.07 & $-70^\circ$~~ & ~~$B_s\to VP$ & 0.90 & $-65^\circ$ \\
 $B\to PV$ & 0.87 & $-30^\circ$~~ & ~~$B_s\to PV$ & 0.85 & $-30^\circ$ \\
 $B\to K^*\rho $ & 0.78 & $-43^\circ$~~ & ~~$B_s\to VV$ & 0.70 & $-55^\circ$ \\
 $B\to K^*\phi $ & 0.65 & $-53^\circ$~~ & & & \\ \hline
 \end{tabular}
 \end{center}
 \end{table}

\subsection{Power corrections to $a_2$}
As pointed out in \cite{CCcp}, while the discrepancies between experiment  and theory in the heavy quark limit for the  rates of penguin-dominated two-body decays of $B$ mesons and direct \CP asymmetries of $\bar B_d\to K^-\pi^+$, $B^-\to K^-\rho^0$ and $\bar B_d\to \pi^+\pi^-$ are resolved by introducing power corrections coming from penguin annihilation, the signs of  direct {\it CP}-violating effects in  $B^-\to K^-\pi^0, B^-\to K^-\eta$ and $\bar B^0\to\pi^0\pi^0$ are flipped to the wrong ones when confronted with experiment. These new $B$-{\it CP} puzzles in QCDF can be explained by the subleading power corrections to the color-suppressed tree amplitudes due to spectator interactions and/or final-state interactions that not only reproduce correct signs for  aforementioned \CP asymmetries but also accommodate the observed $\bar B_d\to \pi^0\pi^0$ and $\rho^0\pi^0$ rates simultaneously.

Following \cite{CCcp}, power corrections to the color-suppressed topology are parametrized as \be \label{eq:a2}
a_2 \to a_2(1+\rho_C e^{i\phi_C}),
\en
with the unknown parameters $\rho_C$ and $\phi_C$ to be inferred from experiment. We shall use $\phi_C\approx -70^\circ$ and $\rho_C\approx 1.3\,,~0.8\,,~0$ for $B\to PP,VP,VV$ decays \cite{CCcp,CCBud}, respectively. This pattern that soft power corrections to $a_2$ are large for $PP$ modes, moderate for $VP$ ones and very small for $VV$ cases is consistent with the observation made in \cite{Kagana2} that soft power correction dominance is much larger for $PP$ than $VP$ and $VV$ final states.
It has been argued that this has to do with the special nature of the pion which is a $q\bar q$ bound state on the one hand and a nearly massless Nambu-Goldstone boson on the other hand \cite{Kagana2}.

\section{$B_s\to PP$ Decays}

Before proceeding to the numerical results of QCDF calculations, we discuss some model-independent flavor symmetry relations in which
many of $B_s\to PP$ decays can be related to $B_d\to PP$ ones by either $U$-spin or SU(3) symmetry. Hence these  relations can be used to cross-check the dynamical calculations.

\subsection{$U$-spin symmetry}
In the limit of $U$-spin symmetry, some of $B_s$ decays can be related to $B_d$ ones. For example,
\be
A(\bar B_s\to K^+\pi^-) &=& V_{ub}^*V_{ud}\la K^+\pi^-|O_d^u|\bar B_s\ra+V_{cb}^*V_{cd}\la K^+\pi^-|O_d^c|\bar B_s\ra, \non \\
A(\bar B_d\to K^-\pi^+) &=& V_{ub}^*V_{us}\la K^-\pi^+|O_s^u|\bar B_d\ra+V_{cb}^*V_{cs}\la K^-\pi^+|O_s^u|\bar B_d\ra,
\en
where the 4-quark operator $O_s$ is for the $b\to q\bar qs$ transition and $O_d$ for the $b\to q\bar q d$ transition. The assumption of
$U$-spin symmetry implies that under $d \leftrightarrow s$ transitions,
\be
\la K^+\pi^-|O_d^u|\bar B_s\ra=\la K^-\pi^+|O_s^u|\bar B_d\ra, \qquad
\la K^+\pi^-|O_d^c|\bar B_s\ra=\la K^-\pi^+|O_s^c|\bar B_d\ra.
\en
Using the relation
\be
{\rm Im}(V_{ub}^*V_{ud}V_{cb}V_{cd}^*)=-{\rm Im}(V_{ub}^*V_{us}V_{cb}V_{cs}^*),
\en
it is straightforward to show that \cite{HeSU3,Lipkin,Gronau}
\be
|A(\bar B_s\to K^+\pi^-)|^2-|A(B_s\to K^-\pi^+)|^2=|A(B_d\to K^+\pi^-)|^2-|A(\bar B_d\to K^-\pi^+)|^2,
\en
and, consequently,
\be \label{eq:Uspin1}
A_{CP}(\bar B_s\to K^+\pi^-)=-A_{CP}(\bar B_d\to K^-\pi^+)\,{\B(\bar B_d\to K^-\pi^+)\over \B(\bar B_s\to K^+\pi^-)}\,{\tau(B_s)\over \tau(B_d)}.
\en
From the current world averages, $A_{CP}(\bar B_d\to K^-\pi^+)=-0.098^{+0.012}_{-0.011}$, $\B(\bar B_d\to K^-\pi^+)=(19.4\pm0.6)\times 10^{-6}$ \cite{HFAG} and the CDF measurement $\B(\bar B_s\to K^+\pi^-)=(5.0\pm1.1)\times 10^{-6}$ \cite{CDF}, it follows that the prediction $A_{CP}(\bar B_s\to K^+\pi^-)\approx 0.37$ under $U$-spin symmetry is in good agreement with the experimental result $0.39\pm0.15\pm0.08$ obtained by CDF \cite{CDF}. Besides $A_{CP}(\bar B_s\to K^+\pi^-)$, CDF has also measured direct \CP violation in the decay $\bar B_d\to K^-\pi^+$ and obtained \cite{Tonelli}
\be
{\Gamma(\bar B_d\to K^-\pi^+)-\Gamma(B_d\to K^+\pi^-)\over
\Gamma(\bar B_s\to K^+\pi^-)-\Gamma(B_s\to K^-\pi^+)}=-0.83\pm0.41\pm0.12,
\en
which is equal to $-1$ under $U$-spin symmetry. Obviously, the experimental measurement is still limited by statistics.

By the same token, we also have the following $U$-spin relations
\be \label{eq:Uspin2}
A_{CP}(\bar B_s\to K^+K^-)&=& -A_{CP}(\bar B_d\to \pi^+\pi^-)\,{\B(\bar B_d\to \pi^+\pi^-)\over \B(\bar B_s\to K^+K^-)}\,{\tau(B_s)\over \tau(B_d)}, \non \\
A_{CP}(\bar B_s\to K^0\bar K^0)&=&-A_{CP}(\bar B_d\to K^0\bar K^0)\,{\B(\bar B_d\to K^0\bar K^0)\over \B(\bar B_s\to K^0\bar K^0)}\,{\tau(B_s)\over \tau(B_d)}, \non \\
A_{CP}(\bar B_s\to K^0\pi^0)&=&-A_{CP}(\bar B_d\to \bar K^0\pi^0)\,{\B(\bar B_d\to \bar K^0\pi^0)\over \B(\bar B_s\to K^0\pi^0)}\,{\tau(B_s)\over \tau(B_d)}, \non \\
A_{CP}(\bar B_s\to \pi^+\pi^-)&=& -A_{CP}(\bar B_d\to K^+K^-)\,{\B(\bar B_d\to K^+K^-)\over \B(\bar B_s\to \pi^+\pi^-)}\,{\tau(B_s)\over \tau(B_d)}.
\en
Unlike the first $U$-spin symmetry relation (\ref{eq:Uspin1}), the above relations  cannot be tested by the present available data. Nevertheless, they can be checked by our dynamical calculations as shown in Sec.IV.C.5.

\subsection{SU(3) symmetry}
There are some cases where two-body decays of $B_d$ and $B_s$ can be related to each other in the limit of SU(3) symmetry provided that some of the annihilation effects can be neglected. Let us consider the decay amplitudes of the following three pairs in QCDF \cite{BN}:
\be
A(\bar B_s\to K^+\pi^-) &=& \sum_{p=u,c} V_{pb}^*V_{pd}A_{K\pi}(\delta _{pu}\alpha_1+\alpha_4^p+\alpha^p_{4,{\rm EW}}+\beta^p_3-{1\over 2}\beta^p_{3,{\rm EW}}),  \\
A(\bar B_d\to \pi^+\pi^-) &=& \sum_{p=u,c} V_{pb}^*V_{pd}A_{\pi\pi}(\delta _{pu}\alpha_1+\alpha_4^p+\alpha^p_{4,{\rm EW}}+\delta_{pu}\beta_1+\beta^p_3-{1\over 2}\beta^p_{3,{\rm EW}}+{1\over 2}\beta^p_{4,{\rm EW}}), \non
\en
and
\be \label{eq:BsKpKm}
A(\bar B_s\to K^+K^-) &=& \sum_{p=u,c} V_{pb}^*V_{ps}\Big[A_{K\bar K}(\delta _{pu}\alpha_1+\alpha_4^p+\alpha^p_{4,{\rm EW}}+\beta^p_3+\beta^p_4-{1\over 2}\beta^p_{3,{\rm EW}}-{1\over 2}\beta^p_{4,{\rm EW}})  \non \\
&+& B_{\bar KK}(\delta_{pu}b^p_1+b^p_4+b^p_{4,{\rm EW}})\Big], \non \\
A(\bar B_d\to K^-\pi^+) &=& \sum_{p=u,c} V_{pb}^*V_{ps}A_{\pi\bar K}(\delta _{pu}\alpha_1+\alpha_4^p+\alpha^p_{4,{\rm EW}}+\beta^p_3-{1\over 2}\beta^p_{3,{\rm EW}}),
\en
and
\be \label{eq:BsK0K0}
A(\bar B_s\to K^0\bar K^0) &=& \sum_{p=u,c} V_{pb}^*V_{ps}\Big[A_{K\bar K}(\alpha_4^p-{1\over 2}\alpha^p_{4,{\rm EW}}+\beta^p_3+\beta^p_4-{1\over 2}\beta^p_{3,{\rm EW}}-{1\over 2}\beta^p_{4,{\rm EW}})  \non \\
&+& B_{\bar KK}(b^p_4-{1\over 2}b^p_{4,{\rm EW}})\Big], \non \\
A(B^-\to \bar K^0\pi^-) &=& \sum_{p=u,c} V_{pb}^*V_{ps}A_{\pi\bar K}(\alpha_4^p-{1\over 2}\alpha^p_{4,{\rm EW}}+\delta _{pu}\beta_2+\beta^p_3+\beta^p_{3,{\rm EW}}),
\en
with $A_{h_1h_2}\equiv X^{(\bar B_sh_1,h_2)}$,
where the expressions of the flavor operators $\alpha_i$ in terms of $a_i$ and the annihilation operators $\beta_i$ in terms of $b_i$ are shown in Eq. (\ref{eq:alphai}). Roughly speaking, $\alpha_1$ is due to the tree topology, $\alpha_4$ comes from the QCD penguin operators $O_4$ and $O_6$, $\alpha^p_{4,{\rm EW}}$ receives contributions from the electroweak operators $O_8$ and $O_{10}$. From the study of hadronic $B_{u,d}$ decays we learn that annihilation effects are negligible in tree-dominated modes and dominated by the $\beta_3$ term in penguin-dominated decays. Hence,  under the approximation of negligible annihilation contributions to tree-dominated decays and keeping only
the dominant penguin annihilations in penguin-dominated decays,
SU(3) symmetry
(or $U$-spin symmetry acting on the spectator quark of the $B$ meson) implies \cite{He,Gronau}
\be
&& A(\bar B_s\to K^+\pi^-) \approx  A(\bar B_d\to \pi^+\pi^-), \quad
A(\bar B_s\to K^+K^-) \approx  A(\bar B_d\to K^-\pi^+), \non \\
&& A(\bar B_s\to K^0\bar K^0) \approx  A(B^-\to \bar K^0\pi^-).
\en
As will be discussed later, it turns out that among the relations
\be \label{eq:SU3}
\B(\bar B_s\to K^+\pi^-) \approx  \B(\bar B_d\to \pi^+\pi^-), && \quad A_{CP}(\bar B_s\to K^+\pi^-) \approx  A_{CP}(\bar B_d\to \pi^+\pi^-), \non \\
\B(\bar B_s\to K^+K^-) \approx  \B(\bar B_d\to K^-\pi^+), &&\quad A_{CP}(\bar B_s\to K^+K^-) \approx  A_{CP}(\bar B_d\to K^-\pi^+), \non \\
\B(\bar B_s\to K^0\bar K^0) \approx  \B(B^-\to \bar K^0\pi^-), &&\quad A_{CP}(\bar B_s\to K^0\bar K^0) \approx  A_{CP}(B^-\to \bar K^0\pi^-),
\en
the first three ones are experimentally fairly satisfied.

{\squeezetable
\begin{table}[tb]
 \caption{$CP$-averaged branching fractions (in units of $10^{-6}$) of $\bar B_s\to PP$ decays obtained in various approaches.
 In the QCDF calculations, the parameters
$\rho_A$ and $\phi_A$ are taken from Table \ref{tab:rhoA},
 $\rho_C=1.3$ and $\phi_C=-70^\circ$. Sources of theoretical uncertainties are discussed in the text. The pQCD predictions to LO and (partial) NLO are taken from \cite{AliBs} and \cite{Xiao}, respectively. For the decays involving an $\eta$ and/or $\eta'$, two different sets of SCET results are quoted from \cite{Zupan}.
}
 \label{tab:PPBr}
\begin{ruledtabular}
{\footnotesize
 \begin{tabular}{l cccccc}
        {Modes} & Class  &  QCDF (this work) & pQCD (LO) &  pQCD (NLO) &     SCET & Expt. \cite{CDFcp,CDF}
          \\  \hline
   ${\overline{B}}^{0}_{s}\to K^{+}\pi^-$                & $T$
    & $5.3^{+0.4+0.4}_{-0.8-0.5}$
                                                          & $7.6^{+3.3}_{-2.5}$
                                                          & $6.3^{+2.6}_{-1.9}$
                                                          &
                                                          $4.9\pm1.2\pm1.3\pm0.3$
                                                          & $5.0\pm1.1$ \\
   ${\overline{B}}^{0}_{s} \to K^{0}{\pi}^{0}$            & $C$
     & $1.7^{+2.5+1.2}_{-0.8-0.5}$
                                                          & $0.16^{+0.12}_{-0.07}$
                                                          & $0.25^{+0.10}_{-0.07}$ & $0.76\pm0.26\pm0.27\pm0.17$

                                                          &          \\
   $\overline B^0_s\to K^0\eta$                           & $C$
     & $0.75^{+1.10+0.51}_{-0.35-0.22}$
                                                          & $0.11^{+0.08}_{-0.11}$
                                                          & $0.08^{+0.03}_{-0.02}$ & $0.80\pm0.48\pm0.29\pm0.18$
                                                          & \\
                                                          &
                                                          & & & & $0.59\pm0.34\pm0.24\pm0.15$
                                                          &\\
   $\overline B^0_s\to K^0\eta^\prime$                    & $C$
      & $2.8^{+2.5+1.1}_{-1.0-0.8}$
                                                          & $0.72^{+0.36}_{-0.24}$
                                                          & $1.87^{+0.45}_{-0.56}$  & $4.5\pm1.5\pm0.4\pm0.5$
                                                          &\\
                                                          &
                                                          & & & & $3.9\pm1.3\pm0.5\pm0.4$
                                                          & \\
   $\overline B^0_s\to K^+ K^-$                           & $P$
   & $25.2^{+12.7+12.5}_{-~7.2-~9.1}$
                                                          & $13.6^{+8.6}_{-5.2}$
                                                          & $15.6^{+5.1}_{-3.9}$ & $18.2\pm6.7\pm1.1\pm0.5$
                                                          & $25.7\pm3.6$ \footnotemark[1] \\
   $\overline B^0_s\to K^0\overline K^0$                  & $P$
     & $26.1^{+13.5+12.9}_{-~8.1-~9.4}$
                                                          & $15.6^{+9.7}_{-6.0}$
                                                          & $18.0^{+4.7}_{-5.9}$  & $17.7\pm6.6\pm0.5\pm0.6$
                                                          &          \\
   $\overline B^0_s\to\eta\eta$                           & $P$
     & $10.9^{+6.3+5.7}_{-4.0-4.2}$
                                                          & $8.0^{+5.4}_{-3.1}$
                                                          & $10.0^{+3.4}_{-2.6}$  & $7.1\pm6.4\pm0.2\pm0.8$
                                                          &\\
                                                          &
                                                          & & & & $6.4\pm6.3\pm0.1\pm0.7$
                                                          & \\
   $\overline B^0_s\to\eta\eta^\prime$                    &$P$
     & $41.2^{+27.3+17.8}_{-12.9-13.1}$
                                                          & $21.0^{+11.7}_{-~7.2}$
                                                          & $34.9^{+11.6}_{-~9.5}$ & $24.0\pm13.6\pm1.4\pm2.7$
                                                          &\\
                                                          &
                                                          & & & & $23.8\pm13.2\pm1.6\pm2.9$&\\
   $\overline B^0_s\to\eta^\prime\eta^\prime$             & $P$
     & $47.9^{+41.6+20.9}_{-17.1-15.3}$
                                                          & $14.0^{+7.0}_{-4.1}$
                                                          & $25.2^{+8.3}_{-6.5}$   & $44.3\pm19.7\pm2.3\pm17.1$
                                                          & \\
                                                          &
                                                          & & & & $49.4\pm20.6\pm8.4\pm16.2$
                                                          & \\
   $\overline B^0_s\to\pi^0\eta$                          & $P_{EW}$
   & $0.05^{+0.03+0.02}_{-0.01-0.01}$
                                                          & $0.05^{+0.02}_{-0.02}$
                                                          & $0.03^{+0.01}_{-0.01}$  & $0.014\pm0.004\pm0.005\pm0.004$
                                                          & \\
                                                          &
                                                          & & & & $0.016\pm0.0007\pm0.005\pm0.006$
                                                          & \\
   $\overline B^0_s\to\pi^0\eta^\prime$                   &$P_{EW}$
      & $0.04^{+0.01+0.01}_{-0.00-0.00}$
                                                          & $0.11^{+0.05}_{-0.03}$
                                                          & $0.08^{+0.03}_{-0.02}$   & $0.006\pm0.003\pm0.002^{+0.064}_{-0.006}$
                                                          & \\
                                                          &
                                                          & & & & $0.038\pm0.013\pm0.016^{+0.260}_{-0.036}$
                                                          &  \\
   $\overline B^0_s\to\pi^+\pi^-$                         & ann
     & $0.26^{+0.00+0.10}_{-0.00-0.09}$
                                                          & $0.57^{+0.18}_{-0.16}$
                                                          & $0.57^{+0.24}_{-0.22}$   &
                                                          & $<1.2$  \\
   $\overline B^0_s\to\pi^0\pi^0$                         & ann
     & $0.13^{+0.0+0.05}_{-0.0-0.05}$
                                                          & $0.28^{+0.09}_{-0.08}$
                                                          & $0.29^{+0.12}_{-0.12}$  &
                                                          &\\
 \end{tabular}}
\footnotetext[1]{ This is the average of the CDF and Belle measurements, $(24.4\pm1.4\pm3.5)\times 10^{-6}$ \cite{Tonelli} and $(38^{+10}_{-~9}\pm7)\times 10^{-6}$ \cite{Belle:KK}, respectively. The old CDF result on $B_s\to K^+K^-$ can be found in \cite{CDFKKold}.}
 \end{ruledtabular}
 \end{table}

\subsection{Numerical results and comparison with other approaches}

 \begin{table}[tb]
 \caption{Same as Table \ref{tab:PPBr} except for the direct $CP$ asymmetries (in \%) in $\bar B_s\to PP$ decays.
 } \label{tab:PPCP}
\begin{ruledtabular}
{\footnotesize
 \begin{tabular}{l c r c r c c c}
 {Modes} & Class
   &   QCDF (this work) & pQCD (LO) & pQCD (NLO) &  SCET & Expt \cite{CDFcp} \\  \hline
   ${\overline{B}}^{0}_{s}\to K^{+}\pi^-$                      & $T$
        & $20.7^{+5.0+3.9}_{-3.0-8.8}$
                                                                &
                                                                $24.1^{+5.6}_{-4.8}$
                                                                & $25.8^{+5.1}_{-6.3}$  & $20\pm17\pm19\pm5$
                                                                & $39\pm15\pm8$ \\
   ${\overline{B}}^{0}_{s} \to K^{0}{\pi}^{0}$ & $C$
     & $36.3^{+17.4+26.6}_{-18.2-24.3}$
                                                                 &
                                                                 $59.4^{+~7.9}_{-12.5}$
                                                                 & $88.0^{+4.8}_{-8.2}$  & $-58\pm39\pm39\pm13$
                                                                 \\
   $\overline B^0_s\to K^0\eta$    & $C$
      & $33.4^{+22.8+25.7}_{-23.8-21.6}$
                                                                 &
                                                                 $56.4^{+8.0}_{-9.3}$
                                                                 & $96.7^{+1.6}_{-2.5}$   & $-56\pm46\pm14\pm6$
                                                                \\
                                                    &            &
                                                                 & & & $61\pm59\pm12\pm8$
                                                                 &\\
   $\overline B^0_s\to K^0\eta^\prime$  &$C$
      & $-49.3^{+6.2+16.0}_{-5.0+13.0}$
                                                                 &
                                                                 $-19.9^{+5.5}_{-5.3}$
                                                                 & $-35.4^{+3.2}_{-2.5}$   & $-14\pm7\pm16\pm2$
                                                                 \\
                                                    &            &
                                                                 & & & $37\pm8\pm14\pm4$
                                                                 & \\
   $\overline B^0_s\to K^+ K^-$  & $P$
       & $-7.7^{+1.6+4.0}_{-1.2-5.1}$
                                                                 &
                                                                 $-23.3^{+5.0}_{-4.6}$
                                                                 & $-15.6^{+1.9}_{-1.6}$  & $-6\pm5\pm6\pm2$
                                                                 \\
   $\overline B^0_s\to K^0\overline K^0$ & $P$
       & $0.40^{+0.04+0.10}_{-0.04-0.04}$
                                                                 &0
                                                                 & $0.4\pm0.1$ & $<10$
                                                                 \\
   $\overline B^0_s\to\eta\eta$   & $P$
      & $-5.0^{+1.5+3.8}_{-2.5-2.8}$
                                                                 &
                                                                 $-0.6^{+0.6}_{-0.5}$
                                                                 & $0.6^{+0.2}_{-0.0}$   & $7.9\pm4.9\pm2.7\pm1.5$
                                                                 \\
                                                    &            &
                                                                 & & & $-1.1\pm5.0\pm3.9\pm1.0$
                                                                 & \\
   $\overline B^0_s\to\eta\eta^\prime$ & $P$
       & $-0.6^{+0.3+0.5}_{-0.4-0.3}$
                                                                 &
                                                                 $-1.3^{+0.1}_{-0.2}$
                                                                 & $-0.2^{+0.1}_{-0.1}$  & $0.04\pm0.14\pm0.39\pm0.43$
                                                                 \\
                                                    &            &
                                                                 & & & $2.7\pm0.9\pm0.8\pm7.6$
                                                                 & \\
   $\overline B^0_s\to\eta^\prime \eta^\prime$ &$P$
       & $3.2^{+0.8+1.0}_{-0.6-1.2}$
                                                                 &
                                                                 $1.9^{+0.4}_{-0.5}$
                                                                 & $1.4^{+0.2}_{-0.2}$  & $0.9\pm0.4\pm0.6\pm1.9$
                                                                 \\
                                                    &            &
                                                                 & & & $-3.7\pm1.0\pm1.2\pm5.6$
                                                                 &\\
   $\overline B^0_s\to\pi^0\eta$         & $P_{EW}$                        & $96.1^{+~1.6+~1.8}_{-14.3-37.1}$
                                                                 &
                                                                 $-0.4^{+0.3}_{-0.3}$
                                                                 & $40.4^{+4.0}_{-7.4}$   &
                                                                 \\
   $\overline B^0_s\to\pi^0\eta^\prime$  & $P_{EW}$
        & $42.9^{+2.3+31.0}_{-8.1-40.9}$
                                                                 &
                                                                 $20.6^{+3.4}_{-2.9}$
                                                                 & $52.5^{+3.2}_{-2.5}$   &
                                                                 \\
   $\overline B^0_s\to\pi^+\pi^-$             & ann                  & 0  & $-1.2^{+1.2}_{-1.3}$
                                                                 & $0.2^{+2.0}_{-1.5}$
                                                                \\
   $\overline B^0_s\to\pi^0\pi^0$          & ann                      & 0 &  $-1.2^{+1.2}_{-1.2}$
                                                                 & $0.2^{+0.1}_{-1.5}$
                                                                 \\
 \end{tabular}}
 \end{ruledtabular}
 \end{table}

We list in Tables \ref{tab:PPBr} and \ref{tab:PPCP} the branching fractions and \CP asymmetries of $\bar B_s\to PP$ decays evaluated in the frameworks of QCD factorization (this work), pQCD to the lowest order (LO) \cite{AliBs} and to the next-to-leading order (NLO) \cite{Xiao} and soft-collinear effective theory (SCET) \cite{Zupan}. For the decays involving an $\eta$ and/or $\eta'$, two different sets of SCET results are quoted from \cite{Zupan}, corresponding to two distinct SCET parameters regarding to the strong phases of the gluonic charming penguin. The expression for the decay amplitudes of $\bar B_s\to PP$ and $VP$ decays in the QCDF approach can be found in the Appendix of \cite{BN}.

The theoretical errors in QCDF calculations
correspond to the uncertainties  due to the variation of (i) the Gegenbauer moments,
the  decay constants, (ii) the heavy-to-light form factors and the strange
quark mass, and (iii) the wave function of the $B$ meson characterized by the
parameter $\lambda_B$, the power corrections due to weak annihilation and hard
spectator interactions described by the parameters $\rho_{A,H}$, $\phi_{A,H}$,
respectively. To obtain the errors shown in Tables
\ref{tab:PPBr}-\ref{tab:VVfL}, we first scan randomly the points in the
allowed ranges of the above nine parameters  and then add errors in quadrature. As noted in passing, we assign an error $\pm0.1$ and $\pm20^\circ$ to the default values of $\rho_A$ and $\phi_A$, respectively, while $\rho_H$ and $\phi_H$ lie in the ranges $0\leq \rho_H\leq 1$ and $0\leq \phi_{H}\leq 2\pi$.
Specifically, the second error in the table is referred  to the
uncertainties caused by the variation of  $\rho_{A,H}$ and $\phi_{A,H}$, where
all other uncertainties are lumped into the first error.  Power corrections beyond the heavy
quark limit generally give the major theoretical uncertainties.
For theoretical uncertainties in pQCD and SCET approaches, the reader is referred to the references cited in the table captions.

\subsubsection{$\bar B_s\to K^+\pi^-,K^0\pi^0,K^0\eta^{(')}$}
As mentioned before, in this work we shall use the form factor $F_0^{B_sK}(0)=0.24$ obtained  by both lattice and pQCD calculations. If a larger $B_s$ to $K$ transition form factor, say, $F_0^{B_sK}(0)=0.31$, is employed, the predicted $\B(\bar B_s\to K^+\pi^-)$ and $\B(\bar B_s\to K^+K^-)$ will be far above the experimental results. \footnote{A larger branching fraction  $\B(\bar B_s\to K^+\pi^-)=(10.2^{+6.0}_{-5.2})\times 10^{-6}$ was obtained in \cite{BN} within the framework of QCDF using the form factor $F_0^{B_sK}(0)=0.31\pm0.05$\,.}
For $F_0^{B_sK}(0)=0.24$, the calculated $\B(\bar B_s\to K^+\pi^-)=(5.3^{+0.4+0.4}_{-0.8-0.5})\times 10^{-6}$ is in good agreement with the measurement $(5.0\pm0.7\pm0.8)\times 10^{-6}$ \cite{CDF}.  Notice that although the same value of $F_0^{B_sK}$ was used in the leading order pQCD calculation, a larger branching fraction of order $7.6\times 10^{-6}$ was obtained (see Table \ref{tab:PPBr}).

A recent detailed analysis in \cite{Chiang} indicates that SU(3) and factorization only remain approximately valid if the branching fraction of $\bar B_s\to K^+\pi^-$ exceeds its current value of $(5.0\pm 1.1)\times 10^{-6}$ by at least 50\% or if the parameter $\xi$ defined by
\be
\xi\equiv {f_K\over f_\pi}\,{F_0^{B\pi}(m_K^2)\over F_0^{B_sK}(m_\pi^2)}\,{m_{B}^2-m_\pi^2\over m_{B_s}^2-m_K^2}
\en
is more than about 1.2\,. The analysis goes as follows. Writing the amplitudes
$A(B^-\to \bar K^0\pi^-) = V_{cs}V_{cb}^*P$ and
$A(\bar B_d\to K^-\pi^+) = V_{us}V_{ub}^*Te^{i\delta}+V_{cs}V_{cb}^*P$,
the measured $B^-\to \bar K^0\pi^-$ rate sets a constraint on the penguin topology $P$. Since $V_{ub}=|V_{ub}|e^{-i\gamma}$, the measurement of $\bar B_d\to K^-\pi^+$ will put a constraint on $T$ as a function of the unitarity angle $\gamma$. Under $U$-spin symmetry, the amplitude $A(\bar B_s\to K^+\pi^-) = V_{ud}V_{ub}^*T'e^{i\delta'}+V_{cd}V_{cb}^*P'$ can be related to the $\bar B_d\to K^-\pi^+$ one by the relations: $T'=T$, $P'=P$ and $\delta'=\delta$. The data of $\bar B_s\to K^+\pi^-$ will be helpful for pinning down the ratio of $P/T$. The analysis of \cite{Chiang} shows that for the value of $\gamma$ to be consistent with other determinations and for the strong phases $\delta$ and $\delta'$ not different much from each other, then either $\B(\bar B_s\to K^+\pi^-)$ is at least 50\% larger than the current measured value or  the parameter $\xi$ is larger than 1.2\,. Our results of $\xi=1.24$ and $\B(\bar B_s\to K^+\pi^-)\approx 5.3\times 10^{-6}$ are thus consistent with the analysis of  \cite{Chiang}.

It is known that the predicted direct \CP violation for $\bar B_d\to K^-\pi^+$ and $\bar B_s\to K^+\pi^-$ modes in naive QCDF is wrong in sign when compared with experiment (see the predictions in \cite{BN}).
This discrepancy together with the rate deficit problem for penguin-dominated decays can be resolved by introducing power corrections coming from penguin annihilation, corresponding to the ``S4 scenario" of \cite{BN}. Using the values given in Table \ref{tab:rhoA} for the parameters $\rho_A$ and $\phi_A$, we obtain
$\acp(\bar B_d\to K^-\pi^+)=-(7.4^{+1.7+4.3}_{-1.5-4.8})\%$
and $\acp(\bar B_s\to K^+\pi^-)=(20.7^{+5.0+3.9}_{-3.0-8.8})\%$, to be compared with the data $-0.098^{+0.012}_{-0.011}$ \cite{HFAG} and  $0.39\pm0.15\pm0.08$ \cite{CDF}, respectively.

The inclusion of soft corrections to the color-suppressed tree topology has two effects: First, it will enhance the rates of $\bar B_s\to K^0\pi^0,K^0\eta$ by a factor of about 2.5 and $\bar B_s\to K^0\eta'$ slightly. Second, it will flip the sign of {\it CP}-violating asymmetries of the former two  modes. For example, $\B(\bar B_s\to K^0\pi^0)$ is enhanced from $0.7\times 10^{-6}$ to $1.7\times 10^{-6}$, while $\acp(\bar B_s\to K^0\pi^0)$ is changed from $-0.214$ to the order of 0.363 (see Tables \ref{tab:PPBr} and \ref{tab:PPCP}). Note that pQCD predictions of branching fractions for the color-suppressed tree-dominated decays $\bar B_s\to K^0\pi^0,K^0\eta^{(')}$ are much smaller than QCDF and SCET. Nevertheless, pQCD results of $\acp$'s for the above three modes agree in signs with QCDF.

We see from Table \ref{tab:PPCP} that SCET predicts a negative sign for $\acp(\bar B_s\to K^0\pi^0)$, contrary to QCDF and pQCD. This deserves a special discussion. The negative sign of $\acp(\bar B_s\to K^0\pi^0)$ has to do with the fact that SCET predicts $\acp(\bar B_d\to \bar K^0\pi^0)=(5\pm4\pm4\pm1)\%$ \cite{Zupan}. From the $U$-spin symmetry relation (\ref{eq:Uspin2}) we learn that \CP asymmetries of $\bar B_s\to K^0\pi^0$ and $\bar B_d\to \bar K^0\pi^0$ are of opposite sign. Although the current world average $\acp(\bar B_d\to\bar K^0\pi^0)=-0.01\pm 0.10$ from
the BaBar and Belle measurements, $-0.13\pm0.13\pm0.03$ \cite{BaBarK0pi0} and $0.14\pm0.13\pm0.06$ \cite{BelleK0pi0} respectively,  is consistent with no \CP violation, there exist several model-independent determinations of this asymmetry: one is the SU(3) relation \cite{Deshpande}
    \be
    \Delta \Gamma(\bar B_d\to\pi^0\pi^0)=-\Delta \Gamma(\bar B_d\to\bar K^0\pi^0),
    \en
    and the other is the approximate sum rule for \CP rate asymmetries \cite{AS98}
\begin{eqnarray} \label{eq:SR}
\Delta\Gamma(\bar B_d\to K^-\pi^+)+\Delta \Gamma(B^-\to\bar K^0\pi^-)\approx 2[\Delta \Gamma(B^-\to K^-\pi^0)+\Delta \Gamma(\bar B_d\to\bar K^0\pi^0)],
\end{eqnarray}
based on isospin symmetry, where
$\Delta \Gamma(B\to K\pi)\equiv \Gamma(\bar B\to\bar K\bar\pi)-\Gamma(B\to K\pi)$. This sum rule allows us to extract $\acp(\bar B_d\to \bar K^0\pi^0)$ in terms of the other three asymmetries in $K^-\pi^+,K^-\pi^0,\bar K^0\pi^-$ modes that have been measured.
From the current data of branching fractions and \CP asymmetries, the above SU(3) relation and {\it CP}-asymmetry sum rule lead to $-0.073^{+0.042}_{-0.041}$ and
$-0.15\pm 0.04$, respectively, for $\acp(\bar B_d\to\bar K^0\pi^0)$. An analysis based on the topological quark diagrams yields a similar result $-0.08\sim -0.12$ \cite{Chiang09}. All these indicate that direct \CP violation  should be negative for $\bar B_d\to\bar K^0\pi^0$ and hence positive for $\bar B_s\to K^0\pi^0$.

\subsubsection{$\bar B_s\to K^+K^-,K^0\bar K^0$}
The penguin-dominated decays $\bar B_s\to K^+K^-,K^0\bar K^0$  have sizable branching fractions of order $25\times 10^{-6}$ in QCDF. The corresponding pQCD and SCET predictions are slightly smaller (Table \ref{tab:PPBr}). \footnote{An early theoretical estimate yielded $\B(\bar B_s\to K^+K^-)=(35\pm7)\times 10^{-6}$ using the measured $B^0\to K^+\pi^-$ branching fraction \cite{Buras}. Based on QCDF and a combination of $U$-spin and isospin arguments, a result of $(20\pm8\pm4\pm2)\times 10^{-6}$ was obtained in \cite{DescotesGenon}.}
From Eqs. (\ref{eq:BsKpKm}) and (\ref{eq:BsK0K0}) we see that $K^+K^-$ and $K^0\bar K^0$ modes differ mainly in the tree contribution $\alpha_1$ and the annihilation term $\beta_1$ induced by the operator $O_1$, both existing in the former but not in the latter. Since these contributions are CKM suppressed relative to the penguin terms, the above two modes should have similar rates but rather distinct \CP asymmetries. Due to the absence of interference between tree and penguin amplitudes, \CP asymmetry is very small in $\bar B_s\to K^0\bar K^0$, less than 1\%. Using the world average of $\acp(\bar B_d\to\pi^+\pi^-)=0.38\pm0.06$, $\B(\bar B_d\to\pi^+\pi^-)=(5.16\pm0.22)\times 10^{-6}$ \cite{HFAG} and $\B(\bar B_s\to K^+K^-)=(25.7\pm3.6)\times 10^{-6}$ \cite{HFAG},
we find from the first $U$-spin relation in Eq. (\ref{eq:Uspin2}) that $\acp(\bar B_s\to K^+K^-)\approx -0.077$ in the $U$-spin limit, which is in excellent agreement with the QCDF prediction. It is very important to measure the direct \CP asymmetry for this mode.

In the pQCD approach, direct \CP violation of $\bar B_s\to K^0\bar K^0$ vanishes to the lower order as there is only one type of CKM matrix element in its decay amplitude, say $V_{tb}V_{ts}^*$ \cite{AliBs}. To the NLO, penguin loop corrections allow other CKM matrix elements enter into the decay amplitude and induce \CP asymmetry \cite{Xiao}. It turns out that the predicted $\acp(\bar B_s\to K^0\bar K^0)$ is very similar in both QCDF and pQCD (to NLO) approaches. It has been argued that the decay $\bar B_s\to K^0\bar K^0$ is a very promising place to look for effects of New Physics through the measurement of its direct \CP violation \cite{Baek,Ciuchini}. For example, it was shown in \cite{Baek} that $\acp(\bar B_s\to K^0\bar K^0)$, which is not more than 1\% in the SM, can be 10 times larger in the presence of SUSY while its rate remains unaffected.

\subsubsection{$\bar B_s\to \eta^{(')}\eta^{(')}$}
The penguin-dominated $\eta^{(')}\eta^{(')}$ modes have sizable rates, especially $B_s\to\eta'\eta'$, the analog of $B\to K\eta'$ in the $B_s$ sector, has the largest branching fraction of order $\sim 50\times 10^{-6}$ in two-body hadronic decays of the $B_s$ meson. The QCDF predictions in \cite{BN} within the S4 scenario are much bigger, $78\times 10^{-6}$ and $66\times 10^{-6}$ respectively for $\eta\eta'$ and $\eta'\eta'$ modes.  This is because Eq. (\ref{eq:FFetaBN}) rather than (\ref{eq:FFeta}) is employed there for describing the $B_s\to \eta^{(')}$ transition form factors.
One of us (CKC) found that the $B_s\to\eta'\eta'$ branching fraction can even reach the level of $1.0\times 10^{-4}$ in the residual final-state scattering model \cite{Chua}.
It is evident from Table \ref{tab:PPBr} that the pQCD approach to lowest order predicts much smaller $\eta^{(')}\eta^{(')}$ rates even though the form factor $F_0^{B_s\eta_s}(0)=0.30$ is used there.  A recent pQCD calculation involving some NLO corrections from vertex corrections, quark loops and chormo-magnetic penguins exhibits some improvements \cite{Xiao}: the branching fractions of $\eta\eta$, $\eta\eta'$ and $\eta'\eta$  are enhanced from 8.0, 21.0 and 14.0  (in units of $10^{-6}$) to 10.0, 34.9 and 25.2, respectively. The gap between pQCD and QCDF is thus improved.
However, the NLO corrections calculated so far in pQCD are still not the complete results as some other pieces of NLO corrections such as hard spectator and annihilation have not been considered. It is important for the pQCD community to carry out the complete NLO calculations.

Since the decays $\bar B_s\to \eta^{(')}\eta^{(')}$ are penguin dominated and their tree amplitudes are color suppressed, their
direct \CP asymmetries are not large.

\subsubsection{$\bar B_s\to \pi\pi$}
The decays $\bar B_s\to \pi\pi$ proceed only through annihilation with the amplitudes \cite{BN}
\be
A_{\bar B_s\to\pi^+\pi^-}\approx \sqrt{2}A_{\bar B_s\to\pi^0\pi^0}\propto 2B_{\pi\pi}b_4^c.
\en
The predicted $\B(\bar B_s\to\pi^+\pi^-)=2.6\times 10^{-7}$ in QCDF is consistent with the current upper limit of $1.2\times 10^{-6}$ \cite{CDF}. Note that in the absence of power corrections i.e. $\rho_A=0$, the branching ratio will become too small, of order $5\times 10^{-8}$.

\subsubsection{$\bar B_s\to \pi^0\eta^{(')}$}
Since the isospin of the final state is $I=1$, the electroweak penguin is the only loop contribution that can contribute to the decays $\bar B_s\to \pi^0\eta^{(')}$, in analog to the $B^-\to \pi^-\pi^0$ transition. However, unlike the latter, the electroweak penguin amplitude in the former gains a CKM enhancement $\lambda_c^{(s)}/\lambda_u^{(s)}$. Indeed,  $P_{\rm EW}$ dominates over $C$ in $\bar B_s\to \pi^0\eta^{(')}$ decays. It is well known that \CP asymmetry of $B^-\to \pi^-\pi^0$ is very small, of order $10^{-3}$. This is ascribed to the fact that the electroweak penguin there is very suppressed with respect to the color-suppressed tree amplitude $C$. On the contrary, \CP violation of  $\bar B_s\to \pi^0\eta^{(')}$ is very sizable due to the dominant $P_{\rm EW}$. From Tables \ref{tab:PPBr} and \ref{tab:PPCP} we see that the approaches of QCDF and pQCD have similar results for the rates of $\bar B_s\to \pi^0\eta^{(')}$ but quite different predictions for $\acp(\bar B_s\to\pi^0\eta')$.

\begin{table}[tb]
 \caption{Direct $CP$ asymmetries (in \%) in $\bar B_s\to PP$ decays via $U$-spin symmetry. Theoretical results of branching fractions and \CP asymmetries for $\bar B_d\to PP$ are taken from \cite{CCBud}.
 } \label{tab:PPUspin}
\begin{ruledtabular}
{\footnotesize
 \begin{tabular}{l c c |l c c c }
 {Modes} & $\B(10^{-6})$
   &  $A_{CP}(\%)$ & Modes  & $A_{CP}(\%)$($U$-spin) & $A_{CP}(\%)({\rm QCDF})$ &  \\  \hline
   ${\overline{B}}^{0}_{d}\to K^{-}\pi^+$                      & $19.3^{+7.9+8.2}_{-4.8-6.2}$
        & $-7.4^{+1.7+4.3}_{-1.5-4.8}$
                                                                & ${\overline{B}}^{0}_{s}\to K^{+}\pi^-$ &
                                                                $25.9$
                                                                & $20.7^{+5.0+3.9}_{-3.0-8.8}$
                                                                \\
    $\overline B^0_d\to \pi^+ \pi^-$
       & $7.0^{+0.4+0.7}_{-0.7-0.7}$                             & $17.0^{+1.3+4.3}_{-1.2-8.7}$                                     & $\overline B^0_s\to K^+ K^-$
                                                                 &
                                                                 $-4.5$
                                                                 &$-7.7^{+1.6+4.0}_{-1.2-5.1}$
                                                                 \\
   ${\overline{B}}^{0}_{d} \to \bar K^{0}{\pi}^{0}$             &$8.6^{+3.8+3.8}_{-2.2-2.9}$
     & $-10.6^{+2.7+5.6}_{-3.8-4.3}$
                                                                 & ${\overline{B}}^{0}_{s} \to K^{0}{\pi}^{0}$
                                                                 &
                                                                 $51.5$
                                                                 & $36.3^{+17.4+26.6}_{-18.2-24.3}$
                                                                 \\

   $\overline B^0_d\to K^0\overline K^0$
       & $2.1^{+1.0+0.8}_{-0.6-0.6}$                       & $-10.0^{+0.7+1.0}_{-0.7-1.9}$                                               & $\overline B^0_s\to K^0\overline K^0$
                                                                 &0.77
                                                                 & $0.40^{+0.04+0.10}_{-0.04-0.04}$
                                                                 \\
   $\overline B^0_d\to K^+K^-$                                   & $0.10^{+0.03+0.03}_{-0.02-0.03}$     &0                                 & $\overline B^0_s\to\pi^+\pi^-$                                &0
                                                                 &0
                                                                \\
 \end{tabular}}
 \end{ruledtabular}
 \end{table}

\subsubsection{Test of $U$-spin and SU(3) symmetries}
There are five $U$-spin relations shown in Eqs. (\ref{eq:Uspin1}) and (\ref{eq:Uspin2}). We have pointed out before that the relation (\ref{eq:Uspin1}) is experimentally verified. For other relations, we are still lack of the measurements of \CP asymmetries. Nevertheless, since the $U$-spin and SU(3) symmetry breaking is already included in QCDF calculations, we can test quantitatively how good the symmetry is. In Table \ref{tab:PPUspin} we show some of direct \CP asymmetries in $B_s$ decays evaluated using the $U$-spin relations Eqs. (\ref{eq:Uspin1}) and (\ref{eq:Uspin2}) and theoretical inputs for the branching fractions of $B_{d,s}\to PP$ decays and \CP asymmetries of $B_d\to PP$. We see that in general $\acp$ obtained by $U$-spin symmetry is consistent with that obtained from direct QCDF calculations. In \cite{AliBs} two parameters
\be
R_3 &\equiv& {|A(B_s\to \pi^+K^-)|^2-|A(\bar B_s\to \pi^-K^+)|^2\over |A(B_s\to \pi^+K^-)|^2+|A(\bar B_s\to \pi^-K^+)|^2}, \non \\
\Delta &\equiv& {A_{CP}(\bar B_d\to \pi^+K^-)\over A_{CP}(\bar B_s\to \pi^-K^+)}+{\B(\bar B_s\to\pi^-K^+)\over \B(\bar B_d\to \pi^+K^-)}\,{\tau(B_d)\over \tau(B_s)},
\en
are defined to quantify the $U$-spin violation through the deviation of $R_3$ from $-1$ and $\Delta$ from 0. However, it is not suitable for the $U$-spin pair ($\bar B_s\to K^0\bar K^0,~\bar B_d\to K^0\bar K^0$) for which we find $\Delta\approx -12$. In this case, it is better to compare $A_{CP}(\bar B_s\to K^0\bar K^0)$ obtained from the $U$-spin relation with the QCDF prediction as we have done in Table \ref{tab:PPUspin}.

As for the test of SU(3) symmetry,  the first three
relations in (\ref{eq:SU3}) are experimentally satisfied:
\be
5.0\pm1.1 \doteq 5.16\pm0.22,\quad 0.39\pm0.17\doteq 0.38\pm0.06, \quad 24.4\pm4.8 \doteq 19.4\pm0.6,
\en
where the branching fractions are in units of $10^{-6}$ and the data are taken from \cite{HFAG}. For the last three relations of (\ref{eq:SU3}) we have
\be
-0.077^{+0.043}_{-0.052}\doteq -0.098^{+0.012}_{-0.011}, \quad
26.1^{+18.7}_{-12.4}\doteq 19.4\pm0.6, \quad 0.004^{+0.001}_{-0.006}\doteq 0.009\pm0.025,
\en
where we have used the theoretical inputs for $B_s$ decays and experimental inputs for $B_d$ ones. Again, it appears that SU(3) symmetry relations are satisfactorily respected.

\begin{table}[tb]
\caption{Same as Table \ref{tab:PPBr} except for the mixing-induced \CP asymmetries $S_f$  in $\bar B_s\to PP$
decays. The parameter $\eta_f=1$ except for $K_S(\pi^0,\eta,\eta')$ modes where $\eta_f=-1$.}  \label{tab:PPmixing}
{\footnotesize
\begin{ruledtabular}
 \begin{tabular}{l c c c c c}
   {Modes} & Class
   &   QCDF (this work) & pQCD (LO) & pQCD (NLO) & SCET  \\  \hline
    $\overline B^0_s\to K_S\pi^0$               &$C$
                                                & $0.08^{+0.29+0.23}_{-0.27-0.26}$
                                                &
                                                $-0.61^{+0.24}_{-0.20}$
                                                & $-0.41^{+0.09}_{-0.13}$ & $-0.16\pm0.41\pm0.33\pm0.17$
                                                \\
    $\overline B^0_s\to K_S\eta$                & $C$
                                                &$0.26^{+0.33+0.21}_{-0.44-0.30}$
                                                &
                                                $-0.43^{+0.23}_{-0.23}$
                                                & $-0.18^{+0.12}_{-0.23}$ &
                                                $0.82\pm 0.32\pm 0.11\pm 0.04$ \\
                                                & & & & &
                                                $0.63\pm 0.61\pm 0.16\pm 0.08$
                                                \\
    $\overline B^0_s\to K_S\eta^\prime$         & $C$
                                                &$0.08^{+0.21+0.20}_{-0.17-0.16}$
                                                &
                                                $-0.68^{+0.06}_{-0.05}$
                                                & $-0.46^{+0.12}_{-0.23}$  &
                                                $0.38\pm 0.08\pm 0.10\pm 0.04$ \\
                                                & & & & &
                                                $0.24\pm 0.09\pm 0.15\pm 0.05$
                                                \\
    $\overline B^0_s\to K^-K^+$                 & $P$
                                                &$0.22^{+0.04+0.05}_{-0.05-0.03}$
                                                &
                                                $0.28^{+0.05}_{-0.05}$
                                                & $0.22^{+0.04}_{-0.03}$  &
                                                $0.19\pm0.04\pm0.04\pm0.01$
                                                \\
    $\overline B^0_s\to K^0\overline K^0$       & $P$
                                                &$0.004^{+0.0+0.002}_{-0.0-0.001}$
                                                &  $0.04$
                                                & $0.04^{+0.00}_{-0.00}$ &
                                                \\
    $\overline B^0_s\to \eta\eta$               & $P$
                                                &$-0.07^{+0.03+0.04}_{-0.06-0.05}$
                                                &
                                                $0.03^{+0.01}_{-0.01}$
                                                & $0.02^{+0.00}_{-0.00}$ &
                                                $-0.026\pm 0.040\pm 0.030\pm0.014$  \\
                                                & & & & &
                                                $-0.077\pm 0.061\pm 0.022\pm 0.026$
                                                \\
    $\overline B^0_s\to \eta\eta'$              & $P$
                                                &$-0.01^{+0.00-0.00}_{-0.01-0.00}$
                                                &
                                                $0.04^{+0.00}_{-0.00}$
                                                & $0.04^{+0.00}_{-0.00}$  &
                                                $0.041\pm 0.004\pm 0.002\pm0.051$ \\
                                                & & & & &
                                                $0.015\pm 0.010\pm 0.008\pm 0.069$
                                                \\
    $\overline B^0_s\to \eta'\eta'$             & $P$
                                                &$0.04^{+0.01+0.01}_{-0.01-0.01}$
                                                &
                                                $0.04^{+0.01}_{-0.01}$
                                                & $0.05^{+0.00}_{-0.01}$  &
                                                $0.049\pm 0.005\pm 0.005\pm0.031$ \\
                                                & & & & &
                                                $0.051\pm 0.009\pm 0.017\pm 0.039$
                                                \\
     $\overline B^0_s\to \pi^0\eta$              & $P_{EW}$
                                                &$0.26^{+0.06+0.48}_{-0.23-0.47}$
                                                &
                                                $0.17^{+0.11}_{-0.13}$
                                                & $0.28^{+0.05}_{-0.05}$  &
                                                $0.45\pm 0.14\pm 0.42\pm 0.30$ \\
                                                & & & & &
                                                $0.38\pm 0.20\pm 0.42\pm 0.37$
                                                \\
    $\overline B^0_s\to \pi^0\eta^\prime$       & $P_{EW}$
                                                &$0.88^{+0.03+0.04}_{-0.15-0.29}$
                                                &
                                                $-0.17^{+0.08}_{-0.09}$
                                                & $-0.18^{+0.12}_{-0.23}$  &
                                                $0.45\pm 0.14\pm 0.42\pm 0.30$ \\
                                                & & & & &
                                                $0.38\pm 0.20\pm 0.42\pm 0.37$
                                                \\
     $\overline B^0_s\to \pi^+\pi^-$            & ann
                                                &$0.15^{+0.00+0}_{-0.00-0}$
                                                &
                                                $0.14^{+0.12}_{-0.06}$
                                                & $0.09^{+0.02}_{-0.00}$  &
                                                \\
     $\overline B^0_s\to \pi^0\pi^0$            &  ann
                                                &$0.15^{+0.00+0}_{-0.00-0}$
                                                &
                                                $0.14^{+0.12}_{-0.06}$
                                                & $0.08^{+0.00}_{-0.00}$  &
                                                \\
      \end{tabular} \end{ruledtabular}
 }
 \end{table}

\subsubsection{Mixing-induced \CP asymmetry}
Measurements of time-dependent \CP
asymmetries in neutral $B_s$ meson decays into a final \CP eigenstate $f$ that is common to $B_s$ and $\bar B_s$ will provide the information on two interesting quantities: mixing-induced \CP asymmetry
 $S_f$   and direct \CP violation
$A_f$ which can be expressed as
 \be
 A_f=-{1-|\lambda_f|^2\over 1+|\lambda_f|^2}, \qquad S_f={2\,{\rm
 Im}\lambda_f\over 1+|\lambda_f|^2},
 \en
where
\be
 \lambda_f={q_{_{B_s}}\over p_{_{B_s}}}\,{A(\ov B_s\to f)\over A(B_s\to f)}={V_{tb}^*V_{ts}\over V_{tb}V_{ts}^*}\,{A(\ov B_s\to f)\over A(B_s\to f)}.
\en
Now let $q_{_{B_s}}/p_{_{B_s}} = e^{2i\beta_s}$ and
\be
\bar A(\bar B_s\to f) &=& A_1e^{i(\phi_{A1}+\delta_1)}+A_2e^{i(\phi_{A2}+\delta_2)}, \non \\
A(B_s\to f) &=& \eta_f\left(A_1e^{i(-\phi_{A1}+\delta_1)}+A_2e^{i(-\phi_{A2}+\delta_2)}\right),
\en
where $CP|f\ra=\eta_f|f\ra$ with $\eta_f=1$ ($-1$) for final $CP$-even (odd) states, $\phi_{A1,A2}$ are weak phases and $\delta_{1,2}$ strong phases. It follows that (see e.g. \cite{Silva})
\be \label{eq:lambda_f}
\lambda_f=\eta_f e^{2i\phi_1}\,{1+re^{i(\phi_1-\phi_2)}e^{i\delta}\over 1+re^{-i(\phi_1-\phi_2)}e^{i\delta}},
\en
with $\phi_{1,2}=\phi_{A1,A2}+\beta_s$, $\delta=\delta_2-\delta_1$ and $r=A_2/A_1$.

For $B_s$ decays, the phase $\beta_s$ due to the $B_s-\overline B_s$ mixing is very small in the SM, of order $1^\circ$. For the decays
$\bar B_s\to K^0\bar K^0,\eta\eta,\eta\eta',\eta'\eta'$ dominated by penguin diagrams (tree contributions to $\eta^{(')}\eta^{(')}$ are color suppressed), $r\simeq 0$ and the phase $\phi_{A1}$ due to $V_{cb}V_{cs}^*$ or $V_{tb}V_{ts}^*$ is also very small. Consequently, $S_f$ are small for penguin-dominated $\bar B_s\to PP$ decays except for $\bar B_s\to K^+K^-$ which receives a tree contribution with $\phi_{A2}=\gamma$. We see from Table \ref{tab:PPmixing} that QCDF, pQCDF and SCET all predict $S_{\bar B_s\to K^+K^-}\approx 0.20$. Recently, both CDF \cite{CDFbetas} and  D0 \cite{D0betas} have reported fits to angular and time distributions of
flavor-tagged $B_s\to J/\psi\phi$ decays which favor a larger value of $\beta_s$ deviated from the SM by 1-2$\sigma$ effects. If this is the case, then mixing-induced \CP violation in $\bar B_s\to K^0\bar K^0,\eta\eta,\eta\eta',\eta'\eta'$ could be sizable. Hence, these modes offer rich possibilities of testing New Physics beyond the SM.

Due to the large magnitude and strong phase of $a_2$ induced from soft power corrections to the color-suppressed tree amplitude, for example, $a_2(K\pi)=0.77 e^{-i 52^\circ}$ (or $0.41 e^{-i11^\circ}$ before corrections), \footnote{In the $B_{u,d}$ systems, $a_2(K\pi)=0.51 e^{-i 58^\circ}$ (or $0.27 e^{-i17^\circ}$ before corrections).}
we find that such corrections will flip the sign of $S_f$ into the positive one for the color-suppressed decays $\bar B_s\to K_S(\pi^0,\eta,\eta')$,
while they are all negative in the pQCD approach.
Recently, it has been claimed that in the pQCD approach there exist uncanceled soft divergences in the $k_T$ factorization for the nonfactorizable $B$ meson decay amplitudes \cite{Li09}. This will enhance the nonfactorizable color-suppressed tree amplitudes. It remains to check if the signs of $S_{\bar B_s\to K_S(\pi^0,\eta,\eta')}$ in pQCD will be flipped again under this ``$a_2$" enhancement.

\section{$B_s\to VP$ Decays}
\subsection{Branching fractions}
The tree-dominated decays $\bar B_s\to K^{*+}\pi^-$ and $\rho^-K^+$ with the amplitudes
\be
A(\bar B_s\to K^{*+}\pi^-) &\approx& A_{K^*\pi}(\delta_{pu}a_1+a_4-r_\chi^\pi a_6), \non \\
A(\bar B_s\to \rho^-K^+) &\approx& A_{K\rho}(\delta_{pu}a_1+a_4+r_\chi^\rho a_6),
\en
have branching fractions of order $10^{-5}$. Since $A_{K^*\pi}\equiv X^{(\bar B_s,K^*\pi)}\approx f_\pi A_0^{B_sK^*}(0)m_{B_s}^2$ and $A_{K\rho}\equiv X^{(\bar B_s\to K\rho)}\approx f_\rho F_0^{B_sK}(0)m_{B_s}^2$ [see Eq. (\ref{eq:X})], it is clear that the $\rho^-K^+$ mode has a rate larger than $K^{*+}\pi^-$ due to the hierarchy of the decay constants $f_\rho\gg f_\pi$. The penguin-dominated $\bar B_s\to VP$ decays such as $K^{*-}K^+$ and $K^{*0}\bar K^0$ have rates smaller than the counterparts in the $PP$ sector as the amplitudes  are proportional to $a_4+r_\chi^{K^*}a_6$ or $a_4-r_\chi^K a_6$ for the former and $a_4+r_\chi^K a_6$ for the latter. Since $a_4$ and $a_6$ are of the same sign and $r_\chi^K >r_\chi^{K^*}$, it is evident that the interference of the penguin terms is constructive for $PP$ and either destructive or less constructive for $VP$. The decay $\bar B_s\to \phi K^0$ is dominated by the $b\to d$ penguin transition and its rate is thus much smaller compared to $b\to s$ dominated $\bar B_s\to K^*K$ decays.

We see from Table \ref{tab:VPBr} that the pQCD predictions for the color-suppressed tree-dominated decays $\bar B_s\to K^{*0}\pi^0,\rho^0K^0,\omega K^0,K^{*0}\eta'$ are one order of magnitude smaller than QCDF and SCET in rates.  For example, $\B(\bar B_s\to \rho^0K^0)$ is predicted to be of order $1.9\times 10^{-6}$ in the approach of QCDF, but it is only about $0.08\times 10^{-6}$ in pQCD. The calculated branching fractions in pQCD for
$K^{*0}\eta$ and some of the penguin-dominated decays e.g. $\bar B_s\to K^{*+}K^-,K^{*0}\bar K^0,\phi K^0,\phi\eta'$ are also much smaller than QCDF. In the following we will comment on the decays $\bar B_s\to\phi\eta^{(')}$. While the QCDF approach leads to
$\B(\bar B_s\to\phi\eta')> \B(\bar B_s\to\phi\eta)$, pQCD and SCET predict very different patterns: $\B(\bar B_s\to\phi\eta)\gg \B(\bar B_s\to\phi\eta')$ in the pQCD approach and it is the other way around  in SCET (see Table \ref{tab:VPBr}).
We should stress that the decay rate of $\bar B_s\to\phi\eta'$ is sensitive to the form factor $A_0^{B_s\phi}(0)$. The decay amplitudes of $\bar B_s\to \phi\eta^{(')}$ are given by
\be
A(\bar B_s\to\phi\eta)&=& \cos\theta A(\bar B_s\to \phi\eta_q)-\sin\theta A(\bar B_s\to\phi\eta_s), \non \\
A(\bar B_s\to\phi\eta')&=& \sin\theta A(\bar B_s\to \phi\eta_q)+\cos\theta A(\bar B_s\to\phi\eta_s),
\en
with
\be
A(\bar B_s\to \phi\eta_s)&=& A_{\phi\eta_s}(\alpha_3^p+\alpha_4^p)+A_{\eta_s\phi}(\alpha_3^p+\alpha_4^p), \non \\
\sqrt{2}A(\bar B_s\to \phi\eta_q)&=& A_{\phi\eta_q}(\delta_{pu}\alpha_2+2\alpha_3^p).
\en
Since $\alpha_4^c(\phi\eta_s)=a_4-r_\chi^{\eta_s}a_6$ and $\alpha_4^c(\eta_s\phi)=a_4+r_\chi^\phi a_6$ are of opposite sign (numerically, $\alpha_4^c(\phi\eta_s)\approx 0.038$ and $\alpha_4^c(\eta_s\phi)\approx -0.033$), there is a cancelation between the two penguin amplitudes of $\bar B_s\to \phi\eta_s$. Note that $\alpha_3^c(\phi\eta_s)$ and $\alpha_3^c(\eta_s\phi)$ also are of opposite sign. It turns out that the sign of $A(\bar B_s\to\phi\eta_s)$ depends on the form factor $A_0^{B_s\phi}(0)$. For $A_0^{B_s\phi}(0)=0.32$ as employed in the present work, $\bar B_s\to \phi\eta_s$ and $\bar B_s\to\phi\eta_q$ will contribute constructively to $\bar B_s\to \phi\eta'$ so that $\B(\bar B_s\to\phi\eta')=2.2\times 10^{-6}$. However, if we use the sum-rule prediction $A_0^{B_s\phi}(0)=0.474$ from Eq. (\ref{eq:FFSR}), then
a near cancelation between $\bar B_s\to \phi\eta_s$ and $\bar B_s\to\phi\eta_q$ occurs in the decays $\bar B_s\to \phi\eta'$, so that its branching fraction, of order $10^{-7}$, becomes very small. Hence, it is very important to measure the branching fractions of $\bar B_s\to \phi\eta^{(')}$ to gain the information on the form factor $A_0^{B_s\phi}$.

One unique feature of the $B_s$ decays is that there exist several modes dominated by electroweak penguins: $\bar B_s\to \pi^0\eta^{(')}, \phi\pi^0, \rho^0\eta^{(')}$ and $\phi\rho^0$. The isospin for the final states of these decays is $I=1$ and hence the electroweak penguin is the only loop contribution that one can have. It dominates over the color-suppressed tree contribution due to the large CKM matrix element associated with the electroweak penguin amplitude. Since a large complex electroweak penguin amplitude due to New Physics is also a possible solution to the $B\to K\pi$ \CP puzzle, it has been advocated that this hypothesis can be tested in the decays $\bar B_s\to\phi\pi^0,\phi\rho^0$ whose rates may get an enhancement by an order of magnitude \cite{Hofer}.

\begin{table}[tb]
\caption{$CP$-averaged branching fractions (in units of $10^{-6}$) of
$\bar B_s\to PV$ decays calculated in various approaches.  The LO pQCD predictions  are taken from \cite{AliBs}, while two different sets of SCET results are quoted from \cite{SCETVP}. }\label{tab:VPBr}
\begin{ruledtabular}
{\footnotesize
 \begin{tabular}{l c r c c c }
 {Modes}  & Class &   QCDF (this work) & PQCD  & SCET 1 & SCET 2  \\  \hline
   $\bar B^0_{s}\to K^{*+}\pi^-$            & $T$& $7.8^{+0.4+0.5}_{-0.7-0.7}$
                 \                           &  $7.6^{+2.9+0.4+0.5}_{-2.2-0.5-0.3}$
                                             & $5.9_{-0.5-0.5}^{+0.5+0.5}$
                                             & $6.6_{-0.1-0.7}^{+0.2+0.7}$\\
   $\bar B^0_{s}\to\rho^-K^{+}$            &  $T$ & $14.7^{+1.4+0.9}_{-1.9-1.3}$
                 \                           & $17.8^{+7.7+1.3+1.1}_{-5.6-1.6-0.9}$
                                             & $7.6_{-0.1-0.8}^{+0.3+0.8}$
                                             & $10.2_{-0.5-0.9}^{+0.4+0.9}$\\
   $\bar B^0_{s}\to K^{*0}{\pi}^{0}$        & $C$& $0.89^{+0.80+0.84}_{-0.34-0.35}$
                 \                           & $0.07^{+0.02+0.04+0.01}_{-0.01-0.02-0.01}$
                                             & $0.90_{-0.01-0.11}^{+0.07+0.10}$
                                             & $1.07_{-0.15-0.09}^{+0.16+0.10}$\\
   $\bar B^0_{s}\to {\rho}^{0}K^{0}$        & $C$ & $1.9^{+2.9+1.4}_{-0.9-0.6}$
                 \                           & $0.08^{+0.02+0.07+0.01}_{-0.02-0.03-0.00}$
                                             & $2.0_{-0.2-0.2}^{+0.2+0.2}$
                                             & $0.81_{-0.02-0.09}^{+0.05+0.08}$\\
   $\bar B^0_s\to \omega K^{0}$               & $C$ & $1.6^{+2.2+1.0}_{-0.7-0.5} $
                 \                           & $0.15^{+0.05+0.07+0.02}_{-0.04-0.03-0.01}$
                                             & $0.90_{-0.01-0.11}^{+0.08+0.10}$
                                             & $1.3_{-0.1-0.1}^{+0.1+0.1}$\\
   $\bar B^0_s\to K^{*-}K^+$                & $P$ & $10.3^{+3.0+4.8}_{-2.2-4.2}$
                 \                           & $6.0^{+1.7+1.7+0.7}_{-1.5-1.2-0.3}$
                                             & $8.4_{-3.4-1.3}^{+4.4+1.6}$
                                             & $9.5_{-2.8-1.1}^{+3.2+1.2}$\\
   $\bar B^0_s\to K^{*+} K^-$                & $P$ & $11.3^{+7.0+8.1}_{-3.5-5.1}$
                 \                           & $4.7^{+1.1+2.5+0.0}_{-0.8-1.4-0.0}$
                                             & $9.8_{-3.7-1.4}^{+4.6+1.7}$
                                             & $10.2_{-3.2-1.2}^{+3.8+1.5}$\\
   $\bar B^0_s\to \bar K^{*0}K^0$          & $P$& $10.5^{+3.4+5.1}_{-2.8-4.5}$
                 \                           & $7.3^{+2.5+2.1+0.0}_{-1.7-1.3-0.0}$
                                             & $7.9_{-3.4-1.3}^{+4.4+1.6}$
                                             & $9.3_{-2.8-1.0}^{+3.2+1.2}$\\
   $\bar B^0_s\to K^{*0}\bar K^0$            & $P$ & $10.1^{+7.5+7.7}_{-3.6-4.8}$
                 \                           & $4.3^{+0.7+2.2+0.0}_{-0.7-1.4-0.0}$
                                             & $8.7_{-3.5-1.4}^{+4.4+1.6}$
                                             & $9.4_{-3.1-1.2}^{+3.7+1.4}$\\
   $\bar B^0_s\to \phi K^0$                   & $P$& $0.6^{+0.5+0.4}_{-0.2-0.3} $
                 \                           & $0.16^{+0.04+0.09+0.02}_{-0.03-0.04-0.01}$
                                             & $0.44_{-0.18-0.07}^{+0.23+0.08}$
                                             & $0.54_{-0.17-0.07}^{+0.21+0.08}$\\
   $\bar B^0_s\to\phi \pi^0$                  & $P_{EW}$ & $0.12^{+0.02+0.04}_{-0.01-0.02} $
                 \                           & $0.16^{+0.06+0.02+0.00}_{-0.05-0.02-0.00}$
                                             & $0.07_{-0.00-0.01}^{+0.00+0.01}$
                                             & $0.09_{-0.00-0.01}^{+0.00+0.01}$ \\
  $\bar B^0_s\to \rho^+\pi^-$                & ann
                                             & $0.02^{+0.00+0.01}_{-0.00-0.01}$
                                             & $0.22^{+0.05+0.04+0.00}_{-0.05-0.06-0.01}$ \\
  $\bar B^0_s\to \rho^-\pi^+$                & ann
                                             & $0.02^{+0.00+0.01}_{-0.00-0.01}$
                                             & $0.24^{+0.05+0.05+0.00}_{-0.05-0.06-0.01}$  \\
  $\bar B^0_s\to \rho^0\pi^0$                & ann
                                             & $0.02^{+0.00+0.01}_{-0.00-0.01}$
                                             & $0.23^{+0.05+0.05+0.00}_{-0.05-0.06-0.01}$ \\ \hline
   $\bar B^0_s\to K^{*0}\eta$                & $C$& $0.56^{+0.33+0.35}_{-0.14-0.17}$
                 \                           & $0.17^{+0.04+0.10+0.03}_{-0.04-0.06-0.01}$
                                             & $1.7_{-0.3-0.1}^{+0.3+0.2}$
                                             & $0.62_{-0.14-0.08}^{+0.14+0.07}$\\
   $\bar B^0_s\to K^{*0}\eta^\prime$         & $C$ & $0.90^{+0.69+0.72}_{-0.30-0.41} $
                 \                           & $0.09^{+0.02+0.03+0.01}_{-0.02-0.02-0.01}$
                                             & $0.64_{-0.26-0.11}^{+0.33+0.11}$
                                             & $0.87_{-0.32-0.08}^{+0.35+0.10}$\\
   $\bar B^0_s\to\phi\eta$                   &  $P$& $1.0^{+1.3+3.0}_{-0.1-1.2} $
                 \                           & $3.6^{+1.5+0.8+0.0}_{-1.0-0.6-0.0}$
                                             & $0.59_{-0.59-0.10}^{+2.02+0.12}$
                                             & $0.94_{-0.97-0.13}^{+1.89+0.16}$\\
   $\bar B^0_s\to\phi\eta^\prime$            & $P$& $2.2^{+4.5+8.3}_{-1.9-2.5}$
                 \                           & $0.19^{+0.06+0.19+0.00}_{-0.01-0.13-0.00}$
                                             & $7.3_{-5.4-1.3}^{+7.7+1.6}$
                                             & $4.3_{-3.6-0.6}^{+5.2+0.7}$\\
   $\bar B^0_s\to\omega\eta$                 & $P,C$ & $0.03^{+0.12+0.06}_{-0.02-0.01} $
                 \                           & $0.04^{+0.03+0.05+0.00}_{-0.01-0.02-0.00}$
                                             & $0.04_{-0.02-0.00}^{+0.04+0.00}$
                                             & $0.007_{-0.002-0.001}^{+0.011+0.001}$\\
   $\bar B^0_s\to\omega\eta^\prime$          & $P,C$& $0.15^{+0.27+0.15}_{-0.08-0.06}$
                 \                           & $0.44^{+0.18+0.15+0.00}_{-0.13-0.14-0.01}$
                                             & $0.001_{-0.000-0.000}^{+0.095+0.000}$
                                             & $0.20_{-0.17-0.02}^{+0.34+0.02}$\\
   $\bar B^0_s\to\rho^0\eta$                 & $P_{EW}$ & $0.10^{+0.02+0.02}_{-0.01-0.01} $
                 \                           & $0.06^{+0.03+0.01+0.00}_{-0.02-0.01-0.00}$
                                             & $0.08_{-0.03-0.01}^{+0.04+0.01}$
                                             & $0.06_{-0.02-0.00}^{+0.03+0.00}$\\
   $\bar B^0_s\to\rho^0\eta^\prime$          & $P_{EW}$& $0.16^{+0.06+0.03}_{-0.02-0.03} $
                 \                           & $0.13^{+0.06+0.02+0.00}_{-0.04-0.02-0.01}$
                                             & $0.003_{-0.000-0.000}^{+0.082+0.000}$
                                             & $0.14_{-0.11-0.01}^{+0.24+0.01}$\\
\end{tabular}}
\end{ruledtabular}
 \end{table}

\begin{table}[tb]
\caption{Same as Table \ref{tab:VPBr} except for the direct $CP$ asymmetries (in \%) in $\bar B_s\to PV$ decays.} \label{tab:VPCP}
\begin{ruledtabular}
{\footnotesize
 \begin{tabular}{l c r c r c c }
   {Modes}   &    Class & QCDF (this work) & PQCD  & SCET 1& SCET 2 \\  \hline
 ${\overline{B}}^{0}_{s}\to K^{*+}\pi^-$                 & $T$& $-24.0^{+1.2+7.7}_{-1.5-3.9}$
                                                         & $-19.0^{+2.5+2.7+0.9}_{-2.6-3.4-1.4}$
                                                            & $-9.9_{-16.7-0.7}^{+17.2+0.9}$
                                                            & $-12.4_{-15.3-1.2}^{+17.5+1.1}$\\
   ${\overline{B}}^{0}_{s}\to\rho^-K^{+}$                 & $T$& $11.7^{+3.5+10.1}_{-2.1-11.6}$
                                                            & $14.2^{+2.4+2.3+1.2}_{-2.2-1.6-0.7}$
                                                            & $11.8_{-20.0-1.1}^{+17.5+1.2}$
                                                            & $10.8_{-10.2-1.0}^{+9.4+0.9}$ \\
   ${\overline{B}}^{0}_{s}\to K^{*0}{\pi}^{0}$             & $C$& $-26.3^{+10.8+42.2}_{-10.9-36.7}$
                                                            & $-47.1^{+7.4+35.5+2.9}_{-8.7-29.8-7.0}$
                                                            & $22.9_{-40.2-1.9}^{+33.1+2.1}$
                                                            & $13.4_{-18.8-1.2}^{+18.6+0.8}$ \\
   ${\overline{B}}^{0}_{s}\to{\rho}^{0}K^{0}$             & $C$& $28.9^{+14.6+25.0}_{-14.5-23.7}$
                                                            & $73.4^{+6.4+16.2+2.2}_{-11.7-47.8-3.9}$
                                                            & $-12.0_{-19.6-0.7}^{+20.1+1.0}$
                                                            & $-32.5_{-23.4-2.9}^{+30.7+2.7}$ \\
   $\overline B^0_s\to \omega K^{0}$                         & $C$& $-32.0^{+18.9+23.6}_{-17.5-26.2} $
                                                            & $-52.1^{+3.2+22.7+3.2}_{-0.0-15.1-2.0}$
                                                            & $24.4_{-41.4-2.0}^{+33.7+2.2}$
                                                            & $18.2_{-17.0-1.7}^{+16.4+1.2}$ \\
   $\overline B^0_s\to  K^{*-}K^+$                          & $P$ & $-11.0^{+0.5+14.0}_{-0.4-18.8}$
                                                            & $-36.6^{+2.3+2.8+1.3}_{-2.3-3.5-1.2}$
                                                            & $-11.2_{-16.2-1.3}^{+19.1+1.3}$
                                                            & $-12.3_{-11.3-0.8}^{+11.4+0.8}$  \\
   $\overline B^0_s\to K^{*+} K^-$                          & $P$& $25.5^{+9.2+16.3}_{-8.8-11.3}$
                                                            &  $55.3^{+4.4+8.5+5.1}_{-4.9-9.8-2.5}$
                                                            & $7.1_{-12.4-0.7}^{+11.2+0.7}$
                                                            & $9.6_{-13.5-0.9}^{+13.0+0.7}$\\
   $\overline B^0_s\to \overline K^{*0}K^{0}$               & $P$& $0.49^{+0.08+0.09}_{-0.07-0.12}$
                                                            &  $0$ &$0$& $0$  \\
   $\overline B^0_s\to K^{*0}\overline K^0$                 & $P$& $0.10^{+0.08+0.05}_{-0.07-0.02}$
                                                            &  $0$  &$0$& $0$\\
   $\overline B^0_s\to \phi K^0$                            & $P$& $-3.2^{+1.2+0.6}_{-1.4-1.3} $
                                                            &  $0$
                                                            & $-3.0_{-4.7-0.3}^{+5.3+0.3}$
                                                            & $-2.2_{-2.9-0.1}^{+3.0+0.1}$\\
   $\overline B^0_s\to \phi\pi^0$                            &$P_{EW}$& $82.2^{+10.9+~9.0}_{-14.0-55.3}$
                                                            & $13.3^{+0.3+2.1+1.5}_{-0.4-1.7-0.7}$
                                                            & $0$&$0$ \\
   $\overline B^0_s\to \rho^+\pi^-$                         & ann
                                                            &$10.2^{+0.8+12.7}_{-0.7-12.8}$
                                                            &$4.6^{+0.0+2.9+0.6}_{-0.6-3.5-0.3}$
                                                            \\
   $\overline B^0_s\to \rho^-\pi^+$                         & ann
                                                            &$-11.1^{+0.7+13.9}_{-0.8-15.7}$
                                                            &$-1.3^{+0.9+2.8+0.1}_{-0.4-3.5-0.2}$
                                                            \\
   $\overline B^0_s\to \rho^0\pi^0$                         & ann
                                                            &$0$
                                                            &$1.7^{+0.2+2.8+0.2}_{-0.8-3.6-0.1}$
                                                            \\ \hline
   $\overline B^0_s\to K^{*0}\eta$                          & $C$& $40.0^{+11.1+53.1}_{-19.2-64.5}$
                                                            & $51.2^{+6.2+14.1+2.0}_{-6.4-12.4-3.3}$
                                                            & $-25.7_{-22.0-1.3}^{+23.4+2.0}$
                                                            & $-62.7_{-22.5-3.9}^{+28.1+2.6}$\\
   $\overline B^0_s\to K^{*0}\eta^\prime$                   & $C$& $-62.5^{+6.0+24.7}_{-5.5-20.2}$
                                                            & $-51.1^{+4.6+15.0+3.2}_{-6.6-18.2-4.1}$
                                                            & $-35.2_{-49.4-3.8}^{+63.3+3.1}$
                                                            & $-32.1_{-23.2-1.7}^{+22.8+2.6}$\\
   $\overline B^0_s\to\phi\eta$                             &$P$&  $-12.4^{+14.1+64.9}_{-~5.7-39.8} $
                                                            &$-1.8^{+0.0+0.6+0.1}_{-0.1-0.6-0.2}$
                                                            & $21.3_{-83.2-2.6}^{+53.5+2.5}$
                                                            & $16.9_{-18.3-1.6}^{+13.8+1.6}$\\
   $\overline B^0_s\to\phi\eta^\prime$                      &$P$& $13.9^{+15.4+28.5}_{-~4.2-89.7}$
                                                            &  $7.8^{+1.5+1.2+0.1}_{-0.5-8.6-0.4}$
                                                            & $4.4_{-7.1-0.6}^{+5.3+0.6}$
                                                            & $7.8_{-4.9-0.8}^{+5.0+0.8}$\\
   $\overline B^0_s\to\omega\eta$                           & $P,C$ & $-64.8^{+24.4+44.0}_{-~3.4-31.6}$                                                                 &  $-16.7^{+5.8+15.4+0.8}_{-3.2-19.1-1.7}$
                                                            &$0$&$0$\\
   $\overline B^0_s\to\omega\eta^\prime$                    & $P,C$& $-39.4^{+4.4+10.4}_{-3.0-11.7}$                           &  $7.7^{+0.4+4.5+9.4}_{-0.1-4.2-0.4}$
                                                            &$0$&$0$\\
   $\overline B^0_s\to\rho^0\eta$                           &  $P_{EW}$& $75.7^{+15.3+13.3}_{-17.6-37.5} $
                                                            &  $-9.2^{+1.0+2.8+0.4}_{-0.4-2.7-0.7}$
                                                            &$0$&$0$ \\
   $\overline B^0_s\to \rho^0\eta^\prime$                    &  $P_{EW}$& $87.4^{+~3.4+~5.7}_{-10.6-30.3} $
                                                            &  $25.8^{+1.3+2.8+3.4}_{-2.0-3.6-1.5}$
                                                            & $0$&$0$\\
\end{tabular}}
\end{ruledtabular}
 \end{table}

\begin{table}[tb]
 \caption{Direct $CP$ asymmetries (in \%) in $B_s\to VP$ decays via $U$-spin symmetry.
 } \label{tab:VPUspin}
\begin{ruledtabular}
{\footnotesize
 \begin{tabular}{l c r |l c r r }
 {Modes} & $\B(10^{-6})$
   &  $A_{CP}(\%)$ & Modes  & $A_{CP}(\%)$($U$-spin) & $A_{CP}(\%)(QCDF)$\\  \hline
   ${\overline{B}}^{0}_{d}\to K^{*-}\pi^+$                      & $9.2^{+1.0+3.7}_{-1.0-3.3}$
        & $-12.1^{+0.5+12.6}_{-0.5-16.0}$
                                                                & ${\overline{B}}^{0}_{s}\to \rho^- K^{+}$ &
                                                                $9.1$
                                                                & $11.7^{+3.5+10.1}_{-2.1-11.6}$
                                                                \\
   ${\overline{B}}^{0}_{d}\to \rho^+ K^{-}$                      & $8.6^{+5.7+7.4}_{-2.8-4.5}$
        & $31.9^{+11.5+19.6}_{-11.0-12.7}$
                                                                & ${\overline{B}}^{0}_{s}\to K^{*+}\pi^-$ &
                                                                $-39.6$
                                                                & $-24.0^{+1.2+7.7}_{-1.5-3.9}$
                                                                \\
     $\overline B^0_d\to K^{*+}K^-$                                   & $0.08^{+0.01+0.02}_{-0.01-0.02}$     & $-4.7^{+0.1+4.7}_{-0.2-2.7}$
                                                                 & $\overline B^0_s\to\rho^+\pi^-$                                &18.9
                                                                 &$10.2^{+0.8+12.7}_{-0.7-12.8}$ \\
    $\overline B^0_d\to K^{*-} K^{+}$                                   & $0.07^{+0.01+0.04}_{-0.01-0.03}$     & $5.5^{+0.2+7.0}_{-0.2-5.5}$
                                                                 & $\overline B^0_s\to\rho^-\pi^+$                                & $-20.5$
                                                                 &$-11.1^{+0.7+13.9}_{-0.8-15.7}$ \\
    $\overline B^0_d\to K^{*0}\overline K^0$
       & $0.70^{+0.18+0.28}_{-0.15-0.25}$                       & $-13.5^{+1.6+1.4}_{-1.7-2.3}$
                                                                 & $\overline B^0_s\to \bar K^{*0}K^0$
                                                                 &0.86
                                                                 & $0.49^{+0.08+0.09}_{-0.07-0.12}$
                                                                 \\
  $\overline B^0_d\to \bar K^{*0}K^{0}$
       & $0.47^{+0.36+0.43}_{-0.17-0.27}$                       & $-3.5^{+1.3+0.7}_{-1.7-2.0}$
                                                                 & $\overline B^0_s\to K^{*0}\overline K^{0}$
                                                                 &0.17
                                                                 & $0.10^{+0.08+0.05}_{-0.07-0.02}$
                                                                 \\
   $\overline B^0_d\to \rho^+ \pi^-$
       & $9.2^{+0.4+0.5}_{-0.7-0.7}$                                           & $-22.7^{+0.9+8.2}_{-1.1-4.4}$                                                  & $\overline B^0_s\to K^{*+} K^-$
                                                                 &
                                                                 $19.0$
                                                                 &$25.5^{+9.2+16.3}_{-8.8-11.3}$
                                                                 \\
   $\overline B^0_d\to \rho^-\pi^+ $
       & $15.9^{+1.1+0.9}_{-1.5-1.1}$                                           & $4.4^{+0.3+5.8}_{-0.3-6.8}$                                                  & $\overline B^0_s\to K^{*-}K^{+} $
                                                                 &
                                                                 $-6.6$
                                                                 &$-11.0^{+0.5+14.0}_{-0.4-18.8}$
                                                                 \\
   \end{tabular}}
 \end{ruledtabular}
 \end{table}

\subsection{Direct \CP asymmetries}
Direct \CP asymmetries of $\bar B_s\to VP$ decays estimated in various approaches are summarized in Table \ref{tab:VPCP}. In QCDF calculations, the signs of \CP asymmetries for color-suppressed tree-dominated decays $\bar B_s\to K^{*0}\pi^0,\rho^0K^0,\omega K^0$ and $K^{*0}\eta$ are governed by the soft corrections to $a_2$ [see Eq. (\ref{eq:a2})].
We see that QCDF and pQCD results agree with each other in signs, whereas SCET predicts opposite signs for these modes. Since the corresponding rates of these decays are very small in pQCD, as a consequence, the {\it CP}-violating asymmetries predicted by pQCD are very large, of order 0.50 or even bigger.

In the pQCD approach, the penguin-dominated decays $\bar B_s\to K^0\phi,\bar K^{*0}K^0,K^{*0}\bar K^0$ have no direct \CP asymmetry as their decay amplitudes are governed by one type of CKM matrix elements, e.g. $V_{tb}V_{td}^*$ for the first mode and $V_{tb}V_{ts}^*$ for the last two. As noticed before for the decay $\bar B_s\to K^0\bar K^0$, NLO corrections from penguin loop interactions can bring a weak phase necessary for a non-vanishing \CP violation. Therefore, it is important to carry out pQCD calculations to NLO for those three modes.
In the approach of SCET, \CP asymmetries of the  decays $\bar B_s\to \pi^0\phi$ and $\bar B_s\to \rho^0(\omega)(\eta,\eta')$ also vanish. As explained in \cite{SCETVP}, there is no charming penguins in these 5 channels and hence no direct \CP violation due to the lack of strong phases.

We use this chance to clarify one misconception about \CP violation under isospin symmetry. The isospin of the final-state  is $I=1$ for $\bar B_s\to \phi\pi^0$, $\rho^0\eta^{(')}$ and $I=0$ for $\bar B_s\to (\phi,\omega)\eta^{(')}$.
One may argue that there is no \CP violation for these decays as they have only one isospin strong phase (see e.g. \cite{Fleischer}). \footnote{
By the same token, it has been (wrongly) claimed that the direct \CP asymmetry is strictly zero in the charged $B^-\to \pi^-\pi^0$ decay.}
On the contrary, we found large direct {\it CP}-violating effects in some of above decays (see Table \ref{tab:VPCP}). The point is that isospin phases should not be confused with other possible strong phases in each of topological amplitudes. In our study, \CP asymmetries of $\bar B_s\to \rho\eta^{(')}$ are large since the electroweak penguins dominate over the color-suppressed tree amplitudes.

\subsection{Test of $U$-spin and SU(3) symmetries}
The pairs related by $U$-spin symmetry are \cite{Suprun}: $(\bar B_d\to K^{*-}\pi^+,~\bar B_s\to \rho^- K^+)$, $(\bar B_d\to K^{-}\rho^+,~\bar B_s\to K^{*+}\pi^-)$, $(\bar B_d\to \rho^{-}\pi^+,~\bar B_s\to K^{*-}K^+)$, $(\bar B_d\to \rho^{+}\pi^-,~\bar B_s\to K^{*+}K^-)$, $(\bar B_d\to K^{*-}K^+,~\bar B_s\to \rho^-\pi^+)$, $(\bar B_d\to K^{*+}K^-,~\bar B_s\to \rho^+\pi^-)$, $(\bar B_d\to \bar K^{*0}K^0,~\bar B_s\to K^{*0}\bar K^0)$, $(\bar B_d\to K^{*0}\bar K^0,~\bar B_s\to \bar K^{*0}K^0)$.  Note that unlike $PP$ and $VV$ modes, $\bar B_s\to K^{*0}\pi^0$ and $\bar B_s\to K^{0}\rho^0$ are not related to $\bar B_d\to \rho^0\bar K^0$ and  $\bar B_d\to \bar K^{*0}\pi^0$, respectively. Direct \CP asymmetries of the pairs listed above are related by $U$-spin symmetry in analogue to Eq. (\ref{eq:Uspin1}) or Eq. (\ref{eq:Uspin2}). The test of $U$-spin symmetry in $B_s\to VP$ decays is shown in Table \ref{tab:VPUspin}. It turns out that $U$-spin symmetry is in general acceptable.

Just as $B_s\to PP$ decays, under the approximation of negligible annihilation contributions to tree-dominated decays and keeping only
the dominant penguin annihilation terms in penguin-dominated decays,
SU(3) symmetry
leads to \cite{He,Gronau}
\be
&& A(\bar B_s\to K^{*+}\pi^-) \approx  A(\bar B_d\to \rho^+\pi^-), \quad  A(\bar B_s\to \rho^- K^{+}) \approx  A(\bar B_d\to \rho^-\pi^+), \non \\
&& A(\bar B_s\to K^{*+}K^-) \approx  A(\bar B_d\to \rho^+ K^-),\quad
A(\bar B_s\to K^{*-}K^+) \approx  A(\bar B_d\to K^{*-}\pi^+).
\en
Thus, we have the relations
\be
&& \B(\bar B_s\to K^{*+}\pi^-) \approx  \B(\bar B_d\to \rho^+\pi^-), \quad  \B(\bar B_s\to \rho^- K^{+}) \approx  \B(\bar B_d\to \rho^-\pi^+), \non \\
&& \B(\bar B_s\to K^{*+}K^-) \approx  \B(\bar B_d\to \rho^+ K^-),\quad
\B(\bar B_s\to K^{*-}K^+) \approx  \B(\bar B_d\to K^{*-}\pi^+),
\en
and
\be
&& A_{CP}(\bar B_s\to K^{*+}\pi^-) \approx  A_{CP}(\bar B_d\to \rho^+\pi^-), \quad  A_{CP}(\bar B_s\to \rho^- K^{+}) \approx  A_{CP}(\bar B_d\to \rho^-\pi^+),  \non \\
&& A_{CP}(\bar B_s\to K^{*+}K^-) \approx  A_{CP}(\bar B_d\to \rho^+ K^-),\quad
A_{CP}(\bar B_s\to K^{*-}K^+) \approx  A_{CP}(\bar B_d\to K^{*-}\pi^+).
\en
Numerically,
\be
 7.8^{+0.6}_{-1.0}\doteq 9.2^{+0.6}_{-1.0}\,, &\quad& 14.7^{+1.7}_{-2.3}\doteq 15.9^{+1.4}_{-1.9}\,, \non \\
 11.3^{+10.7}_{-~6.2}\doteq 8.6^{+9.3}_{-5.3}\,, &\quad& 10.3^{+5.7}_{-4.7}\doteq 9.2^{+3.8}_{-3.4}\,,
\en
for branching fractions in units of $10^{-6}$ and
\be
 -24.0^{+7.8}_{-4.2}\doteq -22.7^{+8.2}_{-4.5}\,, &\quad& 11.7^{+10.7}_{-11.8}\doteq 4.4^{+5.8}_{-6.8}\,, \non \\
 25.5^{+18.7}_{-14.3}\doteq 31.9^{+22.7}_{-16.8}\,, &\quad& -11.0^{+14.0}_{-18.8}\doteq -12.1^{+12.6}_{-16.0}\,,
\en
for direct \CP asymmetries in \%. Hence, the above SU(3) relations are generally respected.

\begin{table}[tb]
\caption{Same as Table \ref{tab:VPBr} except for mixing-induced $CP$ asymmetries $S_f$ in $\bar B_s\to PV$ decays.  The parameter $\eta_f=1$ except for $K_S(\rho^0,\omega,\phi)$ modes where $\eta_f=-1$. Note that the error estimate of $S_{\bar B_s\to K_S\phi}$ is not available in the pQCD calculation  \cite{AliBs}.}
\label{tab:VPmixing}
\begin{ruledtabular}
{\footnotesize
 \begin{tabular}{l c r r r r}
   {Modes}  & Class & QCDF (this work) &     pQCD   &  SCET 1& SCET 2   \\  \hline
 ${\overline{B}}^{0}_{s}\to K_S{\rho}^{0}$ & $C$ & $0.29^{+0.23+0.16}_{-0.24-0.21}$
                                      & $-0.57^{+0.22+0.51+0.02}_{-0.17-0.39-0.05}$
                                      & $0.99_{-0.05-0.01}^{+0.00+0.00}$
                                      & $-0.03^{+0.22+0.17}_{-0.17-0.12}$\\
 $\overline B^0_s\to K_S\omega$       & $C$ & $0.92^{+0.03+0.08}_{-0.07-0.15}$
                                      & $-0.63^{+0.09+0.28+0.01}_{-0.09-0.11-0.02} $
                                      & $-0.11^{+0.28+0.18}_{-0.22-0.14}$
                                      & $0.98^{+0.02+0.00}_{-0.04-0.01}$\\
 $\overline B^0_s\to K_S\phi$         & $P$ & $-0.69^{+0.01+0.01}_{-0.01-0.01}$
                                      & $-0.72$
                                      & $0.09^{+0.04+0.01}_{-0.03-0.01}$
                                      & $-0.13^{+0.02+0.01}_{-0.02-0.01}$\\
 $\overline B^0_s\to\phi\eta$         & $P$ & $0.21^{+0.08+0.61}_{-0.11-0.25}$
                                      & $-0.03^{+0.02+0.07+0.01}_{-0.01-0.20-0.02}$
                                      & $-0.39_{-0.15-0.04}^{+0.43+0.04}$
                                      & $0.23_{-0.16-0.02}^{+0.35+0.02}$\\
 $\overline B^0_s\to\phi\eta^\prime$  & $P$ & $0.08^{+0.05+0.48}_{-0.06-0.81}$
                                      & $0.00^{+0.00+0.02+0.00}_{-0.00-0.02-0.00}$
                                      & $-0.07_{-0.06-0.01}^{+0.06+0.01}$
                                      & $0.10_{-0.05-0.01}^{+0.07+0.01}$\\
 $\overline B^0_s\to\omega\eta$       & $P,C$ & $-0.76^{+0.16+0.52}_{-0.03-0.22}$
                                      & $-0.02^{+0.01+0.02+0.00}_{-0.03-0.08-0.00}$
                                      & $-0.62_{-0.18-0.12}^{+0.41+0.08}$
                                      & $0.93_{-0.98-0.04}^{+0.04+0.03}$\\
 $\overline B^0_s\to\omega\eta^\prime$& $P,C$ & $-0.84^{+0.06+0.04}_{-0.05-0.03}$
                                      & $-0.11^{+0.01+0.04+0.02}_{-0.00-0.04-0.03}$
                                      & $-0.25_{-0.74-0.16}^{+1.23+0.10}$
                                      & $-1.00_{-0.00-0.00}^{+0.04+0.01}$\\
 $\overline B^0_s\to\pi^0\phi$        & $P_{EW}$& $0.40^{+0.04+0.32}_{-0.10-0.53}$
                                      & $-0.07^{+0.01+0.08+0.02}_{-0.01-0.09-0.03}$
                                      & $0.89_{-0.00-0.05}^{+0.00+0.04}$
                                      & $0.90_{-0.00-0.03}^{+0.00+0.02}$\\
 $\overline B^0_s\to\rho^0\eta$       & $P_{EW}$& $0.35^{+0.09+0.22}_{-0.16-0.40}$
                                      & $0.15^{+0.06+0.14+0.01}_{-0.06-0.16-0.01}$
                                      & $1.00_{-0.06-0.01}^{+0.00+0.00}$
                                      & $0.60_{-0.53-0.03}^{+0.30+0.03}$\\
 $\overline B^0_s\to\rho^0\eta^\prime$ & $P_{EW}$& $0.45^{+0.05+0.30}_{-0.13-0.35}$
                                      & $-0.16^{+0.00+0.10+0.04}_{-0.00-0.12-0.05}$
                                      & $0.95_{-1.60-0.02}^{+0.00+0.02}$
                                      & $-0.41_{-0.75-0.15}^{+0.75+0.10}$\\
 $\overline B^0_s\to \rho^0\pi^0$     & ann
                                      & $-0.65^{+0.03+0.00}_{-0.03-0.00}$
                                      & $-0.19^{+0.00+0.02+0.01}_{-0.00-0.02-0.02}$\\
 \end{tabular}}
\end{ruledtabular}
 \end{table}

\subsection{Mixing-induced \CP asymmetry}
As discussed before, due to the tiny phase in the $B_s-\bar B_s$ mixing and in the CKM matrix element $V_{cb}V_{cs}^*$ or $V_{tb}V_{ts}^*$, mixing-induced \CP violation $S_f$ is expected to be very small in the penguin-dominated $\bar B_s\to\phi\eta'$ decays. This is indeed borne out in all model calculations.  The $b\to dg$ penguin-dominated decay $\bar B_s\to K_S\phi$ has a large mixing-induced \CP asymmetry due to the fact that the CKM matrix element $V_{ub}V_{ud}^*$ has a weak phase $-\gamma$. More specifically,
\be \label{eq:ampKphi}
A(\bar B_s\to K_S\phi)\propto V_{ub}V_{ud}^*[A_{K\phi}\alpha_3^u+A_{\phi K}(\alpha_4^u+\beta_3^u)]+V_{cb}V_{cd}^*[A_{K\phi}\alpha_3^c+A_{\phi K}(\alpha_4^c+\beta_3^c)].
\en
To the approximation that $\alpha_{3,4}^c\approx \alpha_{3,4}^u$ and $\beta_3^c\approx \beta_3^u$, it is clear that $A(\bar B_s\to K_S\phi)\propto V_{tb}V_{td}^*=|V_{tb}V_{td}^*|e^{i\beta}$. Thus, $S_{\bar B_s\to K_S\phi}\approx -\sin 2(\beta_s+\beta)=-0.71$ for $\beta_s\approx 1^\circ$ and $\beta=21.58^\circ$ \cite{CKMfitter}. In the pQCD approach, this decay is dominated by the $(S-P)(S+P)$ penguin annihilation process with the CKM matrix element proportional to $V_{tb}V_{td}^*$. Therefore,
both QCDF and pQCD predict $S_{\bar B_s\to K_S\phi}\sim {\cal O}(0.70)$.
(However, no error estimate is done in the pQCD calculation \cite{AliBs}.)
On the contrary, the SCET result of $S_{\bar B_s\to K_S\phi}\sim 0.09$ or $-0.13$ is dramatically different from the QCDF and pQCD predictions. As explained in \cite{SCETVP}, charming penguin contributions to $\bar B_s\to K_S\phi$ dominates over penguin operators and
the CKM matrix element associated with charming penguins is $V_{cb}V_{cd}^*$. Hence, $S_{\bar B_s\to K_S\phi}=-\sin 2\beta_s=-0.03$ is predicted by SCET when penguin contributions are neglected. It should be stressed that although both QCDF and pQCD approaches have similar results for $S_{\bar B_s\to K_S\phi}$, they differ in the prediction of $\acp(\bar B_s\to K_S\phi)$: it is of order $-0.03$ in QCDF and vanishes in pQCD for reasons mentioned above.

The study of \CP violation for
$\bar B_s\to K^{*+}K^-$ and $K^{*-}K^+$ is more complicated as $K^{*\pm}K^\mp$ are not  \CP eigenstates. The time-dependent \CP
asymmetries are given by
 \be \label{eq:CPKstK}
 \A(t) &\equiv & {\Gamma(\ov
B_s^0(t)\to K^{*\pm}K^\mp)-\Gamma(B_s^0(t)\to K^{*\pm}K^\mp)\over
\Gamma(\ov
B_s^0(t)\to K^{*\pm}K^\mp)+\Gamma(B_s^0(t)\to K^{*\pm}K^\mp)} \non \\
 &=& (S\pm \Delta S)\sin(\Delta
 m_st)-(C\pm \Delta C)\cos(\Delta m_st),
 \en
where $\Delta m_s$ is the mass difference of the two neutral $B_s$
eigenstates, $S$ is referred to as mixing-induced \CP asymmetry
and $C$ is the direct \CP asymmetry ($C=-\acp$), while $\Delta S$ and $\Delta
C$ are {\it CP}-conserving quantities. In writing the above equation we have neglected the effects of the width difference of the $B_s$ mesons.
Defining
\begin{eqnarray}
 A_{+-} & \equiv & A(B_s^0\to K^{*+}K^-)~,~~~
A_{-+}  \equiv A(B_s^0\to K^{*-}K^+)~,\nonumber\\
\bar{A}_{-+} & \equiv & A(\overline B_s^0\to K^{*-}K^+)~,~~~
\bar{A}_{+-} \equiv A(\overline B_s^0 \to K^{*+}K^-),
\end{eqnarray}
and
 \be
 \lambda_{+-}={q_{_{B_s}}\over p_{_{B_s}}}\,{\bar A_{+-}\over A_{+-}}, \qquad
 \lambda_{-+}={q_{_{B_s}}\over p_{_{B_s}}}\,{\bar A_{-+}\over A_{-+}},
 \en
we have
 \be \label{eq:C}
 C+\Delta C={1-|\lambda_{+-}|^2\over 1+|\lambda_{+-}|^2}=
 {|A_{+-}|^2-|\bar A_{+-}|^2\over |A_{+-}|^2+|\bar A_{+-}|^2}, \quad
 C-\Delta C={1-|\lambda_{-+}|^2\over 1+|\lambda_{-+}|^2}={|A_{-+}|^2-|\bar A_{-+}|^2\over |A_{-+}|^2+|\bar A_{-+}|^2},
 \en
and
 \be
 S+\Delta S\equiv {2\,{\rm Im}\lambda_{+-}\over 1+|\lambda_{+-}|^2}={2\,{\rm Im}(e^{2i\beta_s}\bar A_{+-}A_{+-}^*)\over
 |A_{+-}|^2+|\bar A_{+-}|^2}, \non \\
  S-\Delta S\equiv {2\,{\rm Im}\lambda_{-+}\over 1+|\lambda_{-+}|^2}={2\,{\rm Im}(e^{2i\beta_s}\bar A_{-+}A_{-+}^*)\over
 |A_{-+}|^2+|\bar A_{-+}|^2}.
 \en
Hence we see that $\Delta S$ describes the strong phase difference
between the amplitudes contributing to $B_s^0\to K^{*\pm}K^\mp$ and
$\Delta C$ measures the asymmetry between $\Gamma(B_s^0\to
K^{*+}K^-)+\Gamma(\ov B_s^0\to K^{*-}K^+)$ and $\Gamma(B_s^0\to
K^{*-}K^+)+\Gamma(\ov B_s^0\to K^{*+}K^-)$.

\begin{table}[tb]
\caption{Various \CP-violating parameters in the decays $\bar B_s^0\to K^{*\pm} K^\mp$. SCET results are quoted from \cite{SCETVP}.
} \label{tab:mixingBstoKstarK}
\begin{ruledtabular}
{\footnotesize
\begin{tabular}{c r r r}
  Parameter  &   QCDF (this work) &   SCET 1                               & SCET 2\\ \hline
  $\A_{K^*K}$ & $0.19^{+0.03+0.14}_{-0.04-0.11}$ \\
  $C$   & $-0.08^{+0.04+0.15}_{-0.04-0.14}$     &  $0.02_{-0.11-0.00}^{+0.10+0.00}$             & $0.01_{-0.09-0.00}^{+0.09+0.00}$        \\
  $S$  & $-0.05^{+0.01+0.13}_{-0.01-0.09}$      &  $-0.02_{-0.07-0.01}^{+0.07+0.01}$             & $0.02_{-0.05-0.00}^{+0.05+0.01}$   \\
  $\Delta C$ & $-0.03^{+0.12+0.46}_{-0.14-0.49}$ &  $-0.09_{-0.10-0.01}^{+0.11+0.01}$             & $-0.11_{-0.09-0.01}^{+0.09+0.01}$  \\
  $\Delta S$ & $0.33^{+0.09+0.30}_{-0.10-0.48}$ &  $0.38_{-0.07-0.04}^{+0.07+0.04}$              & $-0.41_{-0.05-0.03}^{+0.05+0.03}$   \\
\end{tabular}}
\end{ruledtabular}
\end{table}

Next consider the time- and flavor-integrated charge asymmetry
 \be \label{eq:chargeA}
 \A_{K^*K}\equiv {|A_{+-}|^2+|\bar A_{+-}|^2-|A_{-+}|^2-|\bar
 A_{-+}|^2\over |A_{+-}|^2+|\bar A_{+-}|^2+|A_{-+}|^2+|\bar
 A_{-+}|^2},
 \en
Then, following \cite{CKMfitter} one can transform the
experimentally motivated \CP parameters $\A_{K^*K}$ and
$C_{K^*K}$ into the physically motivated choices
 \be
 A_{K^{*+}K^-} &\equiv& {|\kappa^{-+}|^2-1\over |\kappa^{-+}|^2+1},
 \qquad  A_{K^{*-}K^+} \equiv {|\kappa^{+-}|^2-1\over |\kappa^{+-}|^2+1},
 \en
with
 \be
 \kappa^{+-}={q_{_{B_s}}\over p_{_{B_s}}}\,{\bar A_{-+}\over A_{+-}}, \qquad
  \kappa^{-+}={q_{_{B_s}}\over p_{_{B_s}}}\,{\bar A_{+-}\over A_{-+}}.
  \en
Hence,
 \be
 A_{K^{*+}K^-} &=& {\Gamma(\ov B_s^0\to K^{*+}K^-)-\Gamma(B_s^0\to
 K^{*-}K^+)\over \Gamma(\ov B_s^0\to K^{*+}K^-)+\Gamma(B_s^0\to
 K^{*-}K^+)}={\A_{K^*K}-C_{K^*K}-\A_{K^*K}\Delta C_{K^*K}\over 1-\Delta
 C_{K^*K}-\A_{K^*K} C_{K^*K}},   \non \\
 A_{K^{*-}K^+} &=& {\Gamma(\ov B_s^0\to K^{*-}K^+)-\Gamma(B_s^0\to
 K^{*+}K^-)\over \Gamma(\ov B_s^0\to K^{*-}K^+)+\Gamma(B_s^0\to
 K^{*+}K^-)}=-{\A_{K^*K}+C_{K^*K}+\A_{K^*K}\Delta C_{K^*K}\over 1+\Delta
 C_{K^*K}+\A_{K^*K} C_{K^*K}}. \non\\
 \en
Note that the quantities $A_{K^{*\pm}K^\mp}$ here correspond to
$A_{K^{*\mp}K^\pm}$ defined in \cite{CKMfitter}. Therefore,
direct \CP asymmetries $A_{K^{*+}K^-}$ and $A_{K^{*-}K^+}$ are
determined from the above two equations. Results for various \CP-violating parameters in the decays $\bar B_s^0\to K^{*\pm} K^\mp$ are shown in Table \ref{tab:mixingBstoKstarK}.

\section{$B_s\to VV$ Decays}

\subsection{Branching fractions}

In two-body decays $B_{u,d}\to PP,VP,VV$, we have the pattern $VV>PV>VP> PP$ for the branching fractions of tree-dominated modes and $PP>PV\sim VV>VP$ for penguin-dominated ones, where the factorizable amplitude for $B\to VP(PV)$ here is given by $\la V(P)|J_\mu|B\ra\la P(V)|J^\mu|0\ra$.
The first hierarchy is due to the difference of decay constants $f_V>f_P$ and the second hierarchy stems from the fact that the penguin amplitudes  are proportional to $a_4+r_\chi^P a_6$, $a_4+r_\chi^{V}a_6$,  $a_4-r_\chi^P a_6$ $a_4+r_\chi^{V}a_6$, respectively, for $B\to PP,PV,VP,VV$ with $r_\chi^P\sim {\cal O}(1)\gg r_\chi^V$. The same is also true in the $B_s$ sector. From Tables \ref{tab:PPBr}, \ref{tab:VPBr} and \ref{tab:VVBr} we find
\be
&& \B(\bar B_s\to \rho^-K^{*+})>\B(\bar B_s\to \rho^-K^{+})>\B(\bar B_s\to \pi^-K^{*+})>\B(\bar B_s\to \pi^-K^{+}), \non \\
&& \B(\bar B_s\to K^+K^{-})>\B(\bar B_s\to K^{*-}K^+)\sim\B(\bar B_s\to K^{*+}K^-)>\B(\bar B_s\to K^{*-}K^{*+}),
\en
for tree- and penguin-dominated $\bar B_s$ decays, respectively.

\begin{table}[t]
\caption{{\it CP}-averaged branching ratios in $\bar B_s\to VV$ decays
(in units of $10^{-6}$)  obtained in various approaches. Presented are the pQCD predictions taken from \cite{AliBs} and the QCDF predictions from this work and from \cite{BRY} denoted by BRY.}
\label{tab:VVBr}
\begin{ruledtabular}
{\footnotesize
\begin{tabular}{l c c c c c}
 Channel & Class & QCDF (this work) & QCDF (BRY) & pQCD & Expt~\cite{CDF:phiphinew,PDG}\\
\hline \hline
$\bar{B}_s\to \rho^-K^{\ast+}$ & $T$ & $21.6^{+1.3+0.9}_{-2.8-1.5}$ & $25.2^{+1.5+4.7}_{-1.7-3.1}$
& $20.9^{+8.2+1.4+1.2}_{-6.2-1.4-1.1}$ &  \\
$\bar{B}_s \to \rho^0 K^{\ast0}$ & $C$&
$1.3^{+2.0+1.7}_{-0.6-0.3}$ &
$1.5^{+1.0+3.1}_{-0.5-1.5}$
& $0.33^{+0.09+0.14+0.00}_{-0.07-0.09-0.01}$ & $<767$\\
$\bar{B}_s \to \omega K^{\ast0}$ & $C$ & $1.1^{+1.5+1.3}_{-0.5-0.3}$ &
$1.2^{+0.7+2.3}_{-0.3-1.1}$
& $0.31^{+0.10+0.12+0.07}_{-0.07-0.06-0.02}$&  \\
$\bar{B}_s \to K^{*-} K^{*+}$&$P$ & $7.6^{+1.0+2.3}_{-1.0-1.8}$ & $9.1^{+2.5+10.2}_{-2.2-~5.9}$
& $6.7^{+1.5+3.4+0.5}_{-1.2-1.4-0.2}$& \\
$\bar{B}_s \to K^{*0} \overline{K}^{*0}$ &$P$ & $6.6^{+1.1+1.9}_{-1.4-1.7}$ &
$9.1^{+0.5+11.3}_{-0.4-~6.8}$
& $7.8^{+1.9+3.8+0.0}_{-1.5-2.2-0.0}$& $<1681$\\
$\bar{B}_s \to \phi K^{*0}$& $P$ & $0.37^{+0.06+0.24}_{-0.05-0.20}$ & $0.4^{+0.1+0.5}_{-0.1-0.3}$
& $0.65^{+0.16+0.27+0.10}_{-0.13-0.18-0.04}$&$<1013$\\
$\bar{B}_s \to \phi\phi$ & $P$
& $16.7^{+2.6+11.3}_{-2.1-~8.8}$ &  $21.8^{+1.1+30.4}_{-1.1-17.0}$
& $35.3^{+8.3+16.7+0.0}_{-6.9-10.2-0.0}$& $24.0\pm8.9$ \\
$\bar{B}_s \to \phi\omega$ & $P,C$ & $0.18^{+0.44+0.47}_{-0.12-0.04}$ & $0.10^{+0.05+0.48}_{-0.03-0.12}$
& $0.16^{+0.09+0.10+0.01}_{-0.05-0.04-0.00}$&\\
$\bar{B}_s\to\phi\rho^0$& $P_{EW}$ & $0.18^{+0.01+0.09}_{-0.01-0.04}$ & $0.40^{+0.12+0.25}_{-0.10-0.04}$
& $0.23^{+0.09+0.03+0.00}_{-0.07-0.01-0.01}$&$<617$\\
$\bar{B}_s \to \rho^+\rho^-$ &ann& $0.68^{+0.04+0.73}_{-0.04-0.53}$ & $0.34^{+0.03+0.60}_{-0.03-0.38}$
& $1.0^{+0.2+0.3+0.0}_{-0.2-0.2-0.0}$& \\
$\bar{B}_s \to \rho^0\rho^0$ &ann& $0.34^{+0.02+0.36}_{-0.02-0.26}$ & $0.17^{+0.01+0.30}_{-0.01-0.19}$
& $0.51^{+0.12+0.17+0.01}_{-0.11-0.10-0.01}$& $<320$\\
$\bar{B}_s \to \rho^0\omega$ &ann& $0.004^{+0.0+0.005}_{-0.0-0.003}$ &  $<0.01  $
& $0.007^{+0.002+0.001+0.000}_{-0.001-0.001-0.000}$&\\
$\bar{B}_s\to \omega\omega$ &ann& $0.19^{+0.02+0.21}_{-0.02-0.15}$ & $0.11^{+0.01+0.20}_{-0.01-0.12}$
& $0.39^{+0.09+0.13+0.01}_{-0.08-0.07-0.00}$&\\
\end{tabular}}
\end{ruledtabular}
\end{table}

There exist two QCDF calculations of $\bar B_s\to VV$ \cite{BRY,Bartsch}. However, only the longitudinal polarization states of $\bar B_s\to VV$ were considered in \cite{Bartsch}. The analysis in this work differs from Beneke, Rohrer and Yang (BRY) \cite{BRY} mainly in three places: (i) the choice of form factors, (ii) the values of the parameters $\rho_A$ and $\phi_A$, and (iii) the treatment of penguin annihilation contributions characterized by the parameters $\beta_i$ [see Eq. (\ref{eq:beta})] for penguin-dominated $VV$ modes.  First, the form factors for $B_s\to K^*$ and $B_s\to\phi$ transitions we employ in Eq. (\ref{eq:FFour}) are smaller than the ones (\ref{eq:FFBRY}) used by BRY. Second, BRY applied the
values $\rho_A(K^*\phi)=0.6$ and $\phi_A(K^*\phi)=-40^\circ$ obtained from a fit to the data of
$B\to K^*\phi$ to study $B\to\bar K^*\rho$ and $\bar B_s\to VV$ decays. However, as pointed out in \cite{ChengVV}, the parameters  $\rho_A(K^*\rho)\approx 0.78$ and
$\phi_A(K^*\rho)\approx -43^\circ$ fit to the data of $B\to K^*\rho$ decays are slightly different from the ones $\rho_A(K^*\phi)$ and $\phi_A(K^*\phi)$ . Therefore, within the framework of QCDF,
one cannot account for all charmless $B\to VV$ data by a universal set of $\rho_A$ and $\phi_A$ parameters. This explains why the $B\to K^*\rho$ branching fractions obtained by BRY are systematically below the measurements. In this work, we choose $\rho_A=0.70$ and $\phi_A=-55^\circ$ (cf. Table \ref{tab:rhoA}) to describe $B_s\to VV$ decays. Third, as noticed in \cite{ChengVV}, there
are sign errors in the expressions of the annihilation terms $A_3^{f,0}$ and $A_3^{i,0}$ obtained by BRY. As a consequence, BRY claimed (wrongly) that the longitudinal penguin annihilation amplitude $\beta_3^0$ is strongly
suppressed, while the $\beta_3^-$ term receives sizable penguin annihilation contribution. This will affect the decay rates and longitudinal polarization fractions in some of $B\to K^*\rho$ modes, as discussed in details in \cite{ChengVV}. In spite of the above-mentioned three major differences in the calculations of this work and BRY, it turns out that the calculated rates and $f_L$ shown in Tables \ref{tab:VVBr} and \ref{tab:VVfL}, respectively, are similar for most of the $B_s\to VV$ modes.

Recently CDF has reported a new measurement of $B_s\to \phi\phi$ \cite{CDF:phiphinew}
\be
{\B(\bar B_s\to \phi\phi)\over \B(\bar B_s\to J/\psi\phi)}=(1.78\pm0.14\pm0.20)\times 10^{-2}.
\en
Using the branching fraction of $\bar B_s\to J/\psi \phi$ from PDG \cite{PDG}, updated to current values of $f_s/f_d$, this leads to
\be
\B(\bar B_s\to \phi\phi)=(24.0\pm2.1\pm2.7\pm8.2)\times 10^{-6},
\en
where the error is dominated by the last uncertainty coming from the $J/\psi\phi$ branching fraction error. This new measurement is slightly larger than the previous one of $(14^{+8}_{-7})\times 10^{-6}$ \cite{CDFphiphi}. Our prediction $\B(\bar B_s\to\phi\phi)\approx 16.7\times 10^{-6}$ is consistent with experiment.


A few words on the penguin-dominated decays $\bar B_s\to \phi K^{*0}$ and $\bar B_s\to\omega\phi$. Their branching fractions of order $10^{-7}$ are much smaller than other penguin-dominated $K^*\bar K^*$ and $\phi\phi$ modes. This is because $\bar B_s\to \phi K^{*0}$ is induced by the $b\to d$ penguin transition. The
amplitude of $\bar B_s\to\omega\phi$ reads
\be
\sqrt{2}A_{\bar B_s\to \omega\phi}=A_{\phi\omega}\left[\delta_{pu}\alpha_2+2\alpha_3^p+{1\over 2}\alpha_{3,{\rm EW}}^p\right].
\en
The branching fraction due to the QCD penguin $\alpha_3=a_3+a_5$ is small, only at the level of $10^{-7}$. Moreover, there is a partial cancellation between QCD and electroweak penguin contributions, making its rate even smaller.\footnote{It was argued in \cite{BRY} that the color-suppressed tree amplitude $\alpha_2$ is the largest partial amplitude in the decay $\bar B_s\to\omega\phi$. We found that this decay is still dominated by the QCD penguin, though the contribution from $\alpha_2$ is not negligible.}

As seen from Table \ref{tab:VVBr}, pQCD predictions for the color-suppressed tree-dominated modes $\rho^0K^{*0}$ and $\omega K^{*0}$ are much smaller than the QCDF results, whereas $\B(B_s\to\phi\phi)={\cal O}( 35\times 10^{-6})$ is much larger than QCDF and the CDF measurement \cite{CDFphiphi}.

In analog to  Eq. (\ref{eq:SU3}), there are three SU(3) relations relating the rates of $B_s\to VV$ and $B_d\to VV$:
\be
&&\B(\bar B_s\to K^{*+}\rho^-) \approx  \B(\bar B_d\to\rho^+\rho^-), \quad
\B(\bar B_s\to K^{*+}K^{*-}) \approx  \B(\bar B_d\to K^{*-}\rho^+),\non \\
&& \quad \B(\bar B_s\to K^{*0}\bar K^{*0}) \approx  \B(B^-\to \bar K^{*0}\rho^-).
\en
Numerically, we have
\be
21.6^{+1.6}_{-3.2}\doteq 24.2^{+3.1}_{-3.2}, \qquad 7.4^{+2.5}_{-2.1}\doteq 8.9^{+4.9}_{-5.6}, \qquad 6.6\pm2.2\doteq 9.2\pm 1.5
\en
in units of $10^{-6}$, where use of the theoretical calculation of $\B(\bar B_d\to K^{*-}\rho^+)$ from \cite{ChengVV} has been made.

\subsection{Direct \CP violation}

\begin{table}[tb]
\caption{Same as Table \ref{tab:VVBr} except for direct \CP asymmetries (in \%) in the $\bar B_s\to VV$ decays.
}\label{tab:VVCP}
\begin{ruledtabular}
{\footnotesize
\begin{tabular}{l c r r r r}
 Channel & Class & QCDF (this work) & QCDF (BRY) & pQCD\\
\hline
$\bar{B}_s\to \rho^-K^{\ast+}$ & $T$ & $-11^{+1+4}_{-1-1}$ & $-3^{+1+2}_{-1-3}$  &$-8.2^{+1.0+1.2+0.4}_{-1.2-1.7-1.1}$\\
$\bar{B}_s \to \rho^0 K^{\ast0}$ & $C$ & $46^{+15+10}_{-17-25}$ & $27^{+5+34}_{-7-27}$  &$61.8^{+3.2+17.1+4.4}_{-4.7-22.8-2.3}$\\
$\bar{B}_s \to \omega K^{\ast0}$ & $C$ & $-50^{+20+21}_{-15-~6}$ & $-34^{+10+31}_{-~7-43}$ &$-62.1^{+4.8+19.7+5.5}_{-3.9-12.6-1.9}$\\
$\bar{B}_s \to K^{*-} K^{*+}$ & $P$ & $21^{+1+2}_{-2-4}$ & $2^{+0+40}_{-0-15}$ &
$9.3^{+0.4+3.3+0.3}_{-0.7-3.6-0.2}$\\
$\bar{B}_s \to K^{*0} \overline{K}^{*0}$ & $P$ & $0.4^{+0.8+0.6}_{-0.5-0.4}$ & $1^{+0+1}_{-0-0}$ &0\\
$\bar{B}_s \to \phi K^{*0}$ & $P$ & $-9^{+3+4}_{-1-6}$ & $-17^{+4+9}_{-5-9}$ &  0\\
$\bar{B}_s \to \phi\phi$ & $P$ & $0.2^{+0.4+0.5}_{-0.3-0.2}$ & $1^{+0+1}_{-0-0}$ & 0\\
$\bar{B}_s \to \phi\omega$ & $P,C$ & $-8^{+3+20}_{-1-15}$ & $8^{+3+102}_{-3-~56}$ & $3.6^{+0.6+2.4+0.6}_{-0.6-2.4-0.2}$\\
$\bar{B}_s \to \phi\rho^0$ & $P_{EW}$ & $83^{+1+10}_{-0-36}$ & $19^{+5+56}_{-5-67}$
&$10.1^{+0.9+1.6+1.3}_{-0.9-1.8-0.5}$\\
$\bar{B}_s \to \rho^+\rho^-$ & ann & 0 & &$-2.1^{+0.2+1.7+0.1}_{-0.1-1.3-0.1}$\\
$\bar{B}_s \to \rho^0\rho^0$ & ann & 0 & &$-2.1^{+0.2+1.7+0.1}_{-0.1-1.3-0.1}$\\
$\bar{B}_s \to \rho^0\omega$ & ann & 0 & &$6.0^{+0.7+2.7+1.0}_{-0.5-3.9-0.4}$\\
$\bar{B}_s\to \omega\omega$ & ann &  0 & &$-2.0^{+0.1+1.7+0.1}_{-0.1-1.3-0.1}$\\
\end{tabular}\
}
\end{ruledtabular}
 \end{table}

Direct \CP asymmetries in QCDF and pQCD approaches are summarized in Table \ref{tab:VVCP}.

\subsection{Polarization fractions}
For charmless $\ov B\to VV$
decays, it is naively expected that the helicity
amplitudes $\bar \A_h$ (helicities $h=0,-,+$ ) for both tree- and penguin-dominated $\ov B \to VV$ respect the hierarchy pattern
\be \label{eq:hierarchy}
\bar \A_0:\bar \A_-:\bar\A_+=1:\left({\Lambda_{\rm QCD}\over m_b}\right):\left({\Lambda_{\rm QCD}\over
m_b}\right)^2.
\en
Hence,  they are dominated by the longitudinal polarization
states and satisfy the scaling law, namely \cite{Kagan},
 \be \label{eq:scaling}
f_T\equiv 1-f_L={\cal O}\left({m^2_V\over m^2_B}\right), \qquad {f_\bot\over f_\parallel}=1+{\cal
O}\left({m_V\over m_B}\right),
 \en
with $f_L,f_\bot$, $f_\parallel$ and $f_T$ being the longitudinal, perpendicular, parallel and transverse polarization fractions, respectively, defined as
 \be
 f_\alpha\equiv \frac{\Gamma_\alpha}{\Gamma}
                     =\frac{|\bar\A_\alpha|^2}{|\bar\A_0|^2+|\bar\A_\parallel|^2+|\bar\A_\bot|^2},
 \label{eq:f}
 \en
with $\alpha=L,\parallel,\bot$.
In sharp contrast to the
$\rho\rho$ case, the large fraction of transverse polarization of order 0.5 observed in
$\bar B\to \bar K^*\rho$ and $\bar B\to \bar K^*\phi$ decays at $B$ factories  is thus a surprise and poses an interesting challenge for
any theoretical interpretation.  Therefore, in order to obtain a large
transverse polarization in $\bar B\to \bar K^*\rho,\bar K^*\phi$, this scaling law must be
circumvented in one way or another.

As pointed out by Yang and one of us (HYC) \cite{ChengVV}, in the presence of NLO nonfactorizable corrections e.g. vertex,
penguin and hard spectator scattering contributions, effective Wilson coefficients $a_i^h$ are helicity dependent.
Although the factorizable helicity amplitudes $X^0$, $X^-$ and $X^+$ defined by Eq. (\ref{eq:Xh}) respect the scaling law (\ref{eq:hierarchy}) with $\Lambda_{\rm QCD}/m_b$ replaced by $2m_V/m_B$ for the light vector meson production, one needs to consider the effects of helicity-dependent Wilson coefficients: $\A^-/\A^0= f(a_i^-)X^-/[f(a_i^0)X^0]$.
For some penguin-dominated modes, the constructive (destructive) interference in the negative-helicity (longitudinal-helicity) amplitude of the $\ov B\to VV$ decay will render $f(a_i^-)\gg f(a_i^0)$ so that $\A^-$ is comparable to $\A^0$ and the transverse polarization is enhanced. For example, $f_L(\bar K^{*0}\rho^0)\sim 0.91$ is predicted in the absence of NLO corrections. When NLO effects are turned on,  their corrections on $a_i^-$ will render the negative helicity amplitude $\A^-(\bar B^0\to\bar K^{*0}\rho^0)$ comparable to the longitudinal one $\A^0(\bar B^0\to\bar K^{*0}\rho^0)$ so that even at the short-distance level, $f_L$ for $\ov B^0\to \bar K^{*0}\rho^0$ can be as low as 50\%. However, this does not mean that the polarization anomaly is resolved. This is because the calculations based on naive factorization
often predict too small rates for penguin-dominated $\bar B\to VV$ decays, e.g. $\bar B\to \bar K^*\phi$ and $\bar B\to \bar K^*\rho$, by a factor of
$2\sim 3$. Obviously, it does not make sense to compare theory with experiment for $f_{L,T}$ as the definition of polarization fractions depends on the partial rate and hence the prediction can be easily off by a factor of $2\sim 3$. Thus, the first important task is to have some mechanism to bring
up the rates. While the QCD factorization approach
relies on penguin annihilation \cite{Kagan}, soft-collinear effective theory invokes charming
penguin \cite{SCET} and the final-state interaction model considers final-state
rescattering of intermediate charm states \cite{Colangelo,Ladisa,CCSfsi}.
A nice feature of the $(S-P)(S+P)$ penguin annihilation is that it contributes to $\A^0$ and $\A^-$ with similar amount. This together with the NLO corrections will lead to $f_L\sim 0.5$ for penguin-dominated $VV$ modes. Hence, within the framework of QCDF
 we shall assume weak
annihilation to account for the discrepancy between theory and experiment, and fit the existing data of branching fractions and $f_L$ simultaneously by adjusting the
parameters $\rho_A$ and $\phi_A$.
Then using this set of annihilation parameters as a guideline, we can proceed to predict the rates and $f_L$ for other $VV$ decays of the $B_{u,d,s}$ mesons.

The longitudinal polarization fractions in  $\bar B_s\to VV$ decays obtained in the QCDF and pQCD approaches are summarized in Table \ref{tab:VVfL}. Transverse polarization effects are sizable in penguin-dominated $\bar B_s\to VV$ as expected. However, the pQCD calculations indicate that $f_L\sim f_T\sim {1\over 2}$ even for the color-suppressed tree-dominated decays $\bar B_s\to K^{*0}(\rho^0,\omega)$. This is an astonishing result and should be checked by experiment. Polarization fractions of $\bar B_s\to\phi\phi$ will be studied soon by CDF. It will be very interesting to see if the transverse polarization is also important in the penguin dominated $B_s$ decays.

\begin{table}[tb]
\caption{Same as Table \ref{tab:VVBr} except for the longitudinal polarization fractions in the $\bar B_s\to VV$ decays.
}\label{tab:VVfL}
\begin{ruledtabular}
{\footnotesize
\begin{tabular}{l ccccc}
 Channel & Class & QCDF (this work) & QCDF (BRY) & pQCD &
\\
\hline
$\bar{B}_s\to \rho^-K^{\ast+}$ & $T$ &  $0.92^{+0.01+0.01}_{-0.02-0.03}$ & $0.92^{+0.01+0.05}_{-0.01-0.08}$
& $0.937^{+0.001+0.002+0.000}_{-0.002-0.003-0.002}$ & \\
$\bar{B}_s \to \rho^0 K^{\ast0}$ & $C$ & $0.90^{+0.04+0.03}_{-0.05-0.23}$ & $0.93^{+0.02+0.05}_{-0.03-0.54}$
& $0.455^{+0.004+0.069+0.006}_{-0.003-0.043-0.009}$ & \\
$\bar{B}_s \to \omega K^{\ast0}$ & $C$ & $0.90^{+0.03+0.03}_{-0.04-0.23}$ & $0.93^{+0.02+0.05}_{-0.04-0.49}$
& $0.532^{+0.003+0.035+0.023}_{-0.002-0.029-0.013}$ & \\
$\bar{B}_s \to K^{*-} K^{*+}$ &  $P$ & $0.52^{+0.03+0.20}_{-0.05-0.21}$ & $0.67^{+0.04+0.31}_{-0.05-0.26}$
&$0.438^{+0.051+0.021+0.037}_{-0.040-0.023-0.015}$ \\
$\bar{B}_s \to K^{*0} \overline{K}^{*0}$ &  $P$ & $0.56^{+0.04+0.22}_{-0.07-0.26}$ & $0.63^{+0.00+0.42}_{-0.00-0.29}$ &$0.497^{+0.057+0.006+0.000}_{-0.048-0.038-0.000}$
\\
$\bar{B}_s \to \phi K^{*0}$ &  $P$ & $0.43^{+0.02+0.21}_{-0.02-0.18}$ & $0.40^{+0.01+0.67}_{-0.01-0.35}$
&$0.712^{+0.032+0.027+0.000}_{-0.030-0.037-0.000}$
\\
$\bar{B}_s \to \phi\phi$ &  $P$ & $0.36^{+0.03+0.23}_{-0.04-0.18}$ & $0.43^{+0.00+0.01}_{-0.00-0.34}$
& $0.619^{+0.036+0.025+0.000}_{-0.032-0.033-0.000}$ \\
$\bar{B}_s \to \phi\omega$ & $P,C$& $0.95^{+0.01+0.00}_{-0.02-0.42}$ &
&$0.443^{+0.000+0.054+0.009}_{-0.075-0.061-0.004}$
\\
$\bar{B}_s \to \phi\rho^0$ & $P_{EW}$ & $0.88^{+0.01+0.02}_{-0.00-0.18}$ & $0.81^{+0.03+0.09}_{-0.04-0.12}$
&$0.870^{+0.002+0.009+0.009}_{-0.002-0.003-0.004}$ \\
$\bar{B}_s \to \rho^+\rho^-$ & ann & 1 & & $\sim$ 1 \\
$\bar{B}_s \to \rho^0\rho^0$ & ann & 1 & &$\sim$ 1 \\
$\bar{B}_s \to \rho^0\omega$ & ann & 1 & &$\sim$ 1&\\
$\bar{B}_s\to \omega\omega$ & ann & 1 & &$\sim$ 1&\\
\end{tabular}\
}
\end{ruledtabular}
 \end{table}

\begin{table}[tb]
 \caption{Direct $CP$ asymmetries (in \%) in $\bar B_s\to VV$ decays via $U$-spin symmetry.
 } \label{tab:VVUspin}
\begin{ruledtabular}
{\footnotesize
 \begin{tabular}{l r r |l r r r }
 {Modes} & $\B(10^{-6})$
   &  $A_{CP}(\%)$ & Modes  & $A_{CP}(\%)$($U$-spin)~~~~ & $A_{CP}(\%)(QCDF)$ &  \\  \hline
   ${\overline{B}}^{0}_{d}\to K^{*-}\rho^+$                      & $8.9^{+1.1+4.8}_{-1.0-5.5}$
        & $32^{+1+~5}_{-3-24}$~~
                                                                & ${\overline{B}}^{0}_{s}\to K^{*+}\rho^-$ &
                                                                $-10.2$
                                                                & $-11^{+4}_{-1}$
                                                                \\
   ${\overline{B}}^{0}_{d} \to \bar K^{*0}{\rho}^{0}$             &$4.6^{+0.6+3.5}_{-0.5-3.5}$
     & $-15^{+4+16}_{-8-14}$~~
                                                                 & ${\overline{B}}^{0}_{s} \to K^{*0}{\rho}^{0}$
                                                                 &
                                                                 $42.3$
                                                                 & $46^{+18}_{-30}$
                                                                 \\
   $\overline B^0_d\to \rho^+ \rho^-$
       & $25.5^{+1.5+2.4}_{-2.6-1.5}$                             & $-4^{+0+3}_{-0-3}$~~                                     & $\overline B^0_s\to K^{*+} K^{*-}$
                                                                 &
                                                                 $18.7$
                                                                 &$21^{+2}_{-3}$
                                                                 \\
   $\overline B^0_d\to K^{*0}\bar K^{*0}$
       & $0.6^{+0.1+0.2}_{-0.1-0.3}$                       & $-14^{+1+6}_{-1-2}$~~                                               & $\overline B^0_s\to K^{*0}\bar K^{*0}$
                                                                 &0.5
                                                                 & $0.4^{+1.0}_{-0.6}$
                                                                 \\
   $\overline B^0_d\to K^{*+}K^{*-}$                                   & $0.15^{+0.02+0.11}_{-0.01-0.12}$     &0~~                                 & $\overline B^0_s\to\rho^+\rho^-$                                &0
                                                                 &0
                                                                \\
 \end{tabular}}
 \end{ruledtabular}
 \end{table}

\subsection{$U$-spin symmetry}
Analogous to the $\bar B_s\to PP$ sector, $U$-spin symmetry leads to the following relations:
\be \label{eq:UspinVV}
A_{CP}(\bar B_s\to K^{*+}\rho^-)&=& -A_{CP}(\bar B_d\to K^{*-}\rho^+)\,{\B(\bar B_d\to K^{*-}\rho^+)\over \B(\bar B_s\to K^{*+}\rho^-)}\,{\tau(B_s)\over \tau(B_d)}, \non \\
A_{CP}(\bar B_s\to K^{*+}K^{*-})&=& -A_{CP}(\bar B_d\to \rho^+\rho^-)\,{\B(\bar B_d\to \rho^+\rho^-)\over \B(\bar B_s\to K^{*+}K^{*-})}\,{\tau(B_s)\over \tau(B_d)}, \non \\
A_{CP}(\bar B_s\to K^{*0}\bar K^{*0})&=&-A_{CP}(\bar B_d\to K^{*0}\bar K^{*0})\,{\B(\bar B_d\to K^{*0}\bar K^{*0})\over \B(\bar B_s\to K^{*0}\bar K^{*0})}\,{\tau(B_s)\over \tau(B_d)},  \\
A_{CP}(\bar B_s\to K^{*0}\rho^0)&=&-A_{CP}(\bar B_d\to \bar K^{*0}\rho^0)\,{\B(\bar B_d\to \bar K^{*0}\rho^0)\over \B(\bar B_s\to K^{*0}\rho^0)}\,{\tau(B_s)\over \tau(B_d)}, \non \\
A_{CP}(\bar B_s\to \rho^+\rho^-)&=& -A_{CP}(\bar B_d\to K^{*+}K^{*-})\,{\B(\bar B_d\to K^{*+}K^{*-})\over \B(\bar B_s\to \rho^+\rho^-)}\,{\tau(B_s)\over \tau(B_d)}. \non
\en
In Table \ref{tab:VVUspin} we compare the results of \CP asymmetries inferred from $U$-spin relations with the direct QCDF calculations. It appears that $U$-spin symmetry works well in the $VV$ sector.

Assuming  that the transverse amplitude can be expressed as a single dominant contribution which may arise from new physics, $U$-spin symmetry implies that the transverse amplitudes of  $B_s\to VV$ can be related to the $U$-spin related decays in the $B_d$ sector via \cite{NagashimaFPCP}
\be
{\A_T(\bar B_s\to K^{*0}\bar K^{*0})\over \A_T(\bar B_d\to \bar K^{*0}K^{*0})}\approx \left|{V_{ts}\over V_{td}}\right|\,{f_{B_s}\over f_{B_d}}, \qquad
{\A_T(\bar B_d\to \phi \bar K^{*0})\over \A_T(\bar B_d\to \phi \bar K^{*0})}\approx \left|{V_{ts}\over V_{td}}\right|\,{f_{B_d}\over f_{B_s}}.
\en
Therefore,
\be
{f_T(\bar B_s\to K^{*0}\bar K^{*0})\over f_T(\bar B_d\to \bar K^{*0}K^{*0})} &\approx& (25.5\pm6.5) {\B(\bar B_d\to \bar K^{*0}K^{*0})\over \B(\bar B_s\to K^{*0}\bar K^{*0})}, \non \\
{f_T(\bar B_d\to \phi \bar K^{*0})\over f_T(\bar B_s\to  \phi K^{*0})}&\approx& (19.3\pm4.9) {\B(\bar B_s\to \phi K^{*0})\over \B(\bar B_d\to \phi\bar K^{*0})}.
\en
The polarization measurement in the $B_d$ decay thus allows one to predict the transverse polarization in the $B_s$ decay.\footnote{
Based on SU(3) flavor symmetry, it has been shown in \cite{Datta} that the transverse polarizations of $\bar B_s\to\phi\phi$ and $\bar B_s\to \phi K^{*0}$ can be related to
$\bar B_d\to \phi \bar K^{*0}$ and $\bar B_d\to K^{*0}\bar K^{*0}$, respectively.}
Using the data \cite{HFAG}
\be
\B(\bar B_d\to \bar K^{*0}K^{*0})=(1.28^{+0.37}_{-0.32})\times 10^{-6}, &\quad& f_L(\bar B_d\to \bar K^{*0}K^{*0})=0.80^{+0.12}_{-0.13}, \non \\
\B(\bar B_d\to \phi\bar K^{*0})=(9.8\pm0.7)\times 10^{-6}, &\quad& f_L(\bar B_d\to \phi\bar K^{*0})=0.48\pm0.03,
\en
and QCDF predictions for $\B(\bar B_s\to K^{*0}\bar K^{*0})$ and $\B(\bar B_s\to \phi K^{*0})$, we obtain
\be
f_T(\bar B_s\to K^{*0}\bar K^{*0})=1.02\pm 0.28, \qquad
f_T(\bar B_s\to  \phi K^{*0})=0.73\pm 0.19.
\en
It is obvious that the central value of the predicted $f_T(\bar B_s\to K^{*0}\bar K^{*0})$ via $U$-spin symmetry is too large. Note that there is a discrepancy between the QCDF prediction of $\B(\bar B_d\to \bar K^{*0} K^{*0})=(0.6^{+0.2}_{-0.3})\times 10^{-6}$ \cite{ChengVV} and the BaBar measurement $\B(\bar B_d\to \bar K^{*0}K^{*0})=(1.28^{+0.37}_{-0.32})\times 10^{-6}$ \cite{BaBarKstKst}. We need to await a more precise measurement of $\bar B_d\to \bar K^{*0}K^{*0}$ in order to have a more accurate prediction of its transverse polarization fraction via $U$-spin symmetry.

\subsection{Time-dependent \CP violation}
In principle, one can study time-dependent \CP asymmetries
for each helicity component,
 \be
 \A_h(t) &\equiv & {\Gamma(\ov
B_s^0(t)\to V_hV'_h)-\Gamma(B_s^0(t)\to V_hV'_h)\over
\Gamma(\ov
B_s^0(t)\to V_hV'_h)+\Gamma(B_s^0(t)\to V_hV'_h)} \non \\
 &=& S_h\sin(\Delta
 m_st)-C_h\cos(\Delta m_st),
 \en
where the effects of the width difference of the $B_s$ mesons have been neglected. From Table \ref{tab:VVBr} we see that there is only one decay mode of particular interest, namely, $\bar B_s\to \phi\phi$. Indeed, this could be the most promising channel for the forthcoming LHCb experiment.
This channel is a pure $b\to s\bar s s$ penguin-induced process and hence provides an ideal place for exploring the signal of New Physics via $B_s-\bar B_s$ mixing and/or the penguin process. The other decays such as $\bar B_s\to \rho\rho,\rho^0\omega,\omega\omega$ proceed through weak annihilation. The modes $\phi\omega$ and $\phi\rho^0$ receive QCD penguin and electroweak penguin contributions, respectively, but their rates are too small. A straightforward calculation gives
\be
\B_L=(5.9^{+1.0+5.3}_{-0.8-5.7})\times 10^{-6}, \quad C_L=(-0.5^{+0.1+1.4}_{-0.2-1.5})\%, \quad S_L=(-0.5^{+0.1+1.1}_{-0.1-1.8})\%,
\en
for the longitudinal component of $\bar B_s\to\phi\phi$. Note that $S_L$ is found to be positive and small $\leq 0.02$ in \cite{Bartsch}, while our result is negative for $S_L$. An observation of large \CP violation in this decay will rule out the scenario of minimal flavor violation.
Time-dependent \CP violation will be studied at LHC. If LHCb is upgraded to accumulate data sample of 100fb$^{-1}$, the sensitivity of $S_{B_s\to\phi_L\phi_L}$ will reach the level of $0.01\sim 0.02$.

\section{Conclusions}
We have re-examined the branching fractions and {\it CP}-violating asymmetries of charmless $\bar B_s\to PP,~VP,~VV$ decays  in the framework of QCD factorization. We have included subleading power corrections to the penguin annihilation topology and to color-suppressed tree amplitudes that are crucial for resolving the \CP puzzles and rate deficit problems with penguin-dominated two-body decays  and color-suppressed tree-dominated $\pi^0\pi^0$ and $\rho^0\pi^0$ modes in the $B_{u,d}$ sector. Our main results are:

\renewcommand{\theenumi}{\roman{enumi})}
\begin{enumerate}

\item Many model-independent relations for \CP asymmetries and branching fractions of $\bar B_d$ and $\bar B_s$ decays can be derived under $U$-spin and SU(3) symmetries for $PP,VP,VV$ modes. In general, they are either experimentally verified or theoretically satisfied. There are also a few $U$-spin relations for transverse polarizations in $B_s\to VV$ decays.

\item For the $B_s\to K$ transition form factor, we use a smaller one, $F^{B_sK}\approx 0.24$ at $q^2=0$ obtained by the lattice calculation, to avoid too large rates for $\bar B_s\to K^+\pi^-,K^+K^-$ decays.

\item Both QCDF and SCET indicate that the penguin-dominated decay $B_s\to\eta'\eta'$, the analog of $B\to K\eta'$ in the $B_s$ sector, has the largest branching fraction of order $\sim 50\times 10^{-6}$ in two-body hadronic decays of the $B_s$ meson, whereas the pQCD approach claims that $\B(\bar B_s\to\eta\eta')\approx 35\times 10^{-6}$ is the largest one.

\item Even at the decay rate level, there are some noticeable differences between various approaches.
    The branching fractions of the color-suppressed tree-dominated decays obtained by pQCD, for example, $\bar B_s\to K^0\pi^0,K^0\eta^{(')}, K^{*0}\pi^0,\rho^0K^0,\omega K^0,K^{*0}\eta'$ are typically smaller by one order of magnitude than that of QCDF and SCET. For example, $\B(\bar B_s\to \rho^0K^0)$ is predicted to be of order $1.9\times 10^{-6}$ by QCDF, but it is only about $0.08\times 10^{-6}$ in pQCD. In the QCDF approach, many of the above-mentioned decays get a substantial enhancement from the power corrections to the color-suppressed tree topology.

\item
The decay rate of $\bar B_s\to\phi\eta'$ is sensitive to the $B_s\to\phi$ transition form factor $A_0^{B_s\phi}(0)$.
For $A_0^{B_s\phi}(0)=0.474$ obtained by the light-cone sum rule method, a near cancelation between $\bar B_s\to \phi\eta_s$ and $\bar B_s\to\phi\eta_q$ occurs in the decays $\bar B_s\to \phi\eta'$, so that its branching fraction, of order $10^{-7}$, becomes very small. However, if the value $A_0^{B_s\phi}(0)=0.30$ favored by many other model calculations is employed, then $\bar B_s\to \phi\eta_s$ and $\bar B_s\to\phi\eta_q$ will contribute constructively to $\bar B_s\to \phi\eta'$ so that $\B(\bar B_s\to\phi\eta')=2.2\times 10^{-6}$ and $\B(\bar B_s\to\phi\eta)=1.0\times 10^{-6}$. Hence, it is very important to measure the branching fractions of $\bar B_s\to \phi\eta^{(')}$ to gain the information on the form factor $A_0^{B_s\phi}$.

\item Measurements of {\it CP}-violating asymmetries can be used to discriminate between QCDF, pQCD and SCET approaches:

\begin{enumerate}
\item Both QCDF and pQCD predict a positive sign for $\acp(\bar B_s\to K^0\pi^0)$, whereas SCET leads to a negative one. This can be traced back to fact that $\acp(\bar B_d\to \bar K^0\pi^0)$ is positive in SCET, while it is negative inferred from the {\it CP}-asymmetry sum rule, SU(3) relation  and the topological quark diagram analysis.
\item For color-suppressed tree-dominated decays $\bar B_s\to K^{*0}\pi^0,\rho^0K^0,\omega K^0,K^{*0}\eta'$, QCDF and pQCD results are of the same sign, whereas SCET predicts opposite signs for these modes. In the QCDF approach, the signs of these \CP asymmetries  are governed by the soft corrections to $a_2$. Since the corresponding rates of these decays are very small in pQCD, as a consequence, the {\it CP}-violating asymmetries predicted by pQCD are very large, of order 0.50 or even bigger.

\item In the QCDF framework, the penguin-dominated decays $\bar B_s\to K^0\phi,\bar K^{*0}K^0,K^{*0}\bar K^0$
    have non-vanishing \CP asymmetries, though very small for the last two modes, whereas leading order pQCD predicts no \CP violation for these three decays.

\end{enumerate}

\item Mixing-induced \CP asymmetries of the penguin-dominated decays  $\bar B_s\to K^0\bar K^0,\eta^{(')}\eta^{(')},\phi\eta',\phi\phi$ are predicted to be very small in the SM. Especially, we found $S_{\bar B_s\to\phi_L\phi_L}\sim -0.5\%$. They are sensitive to New Physics and provide possibilities of new discoveries. While both QCDF and pQCD approaches predict $S_{\bar B_s\to K_S\phi}\sim {\cal O}(0.70)$,  the SCET result of 0.09 or $-0.13$ is dramatically different.

\item Due to  soft power corrections to the color-suppressed tree amplitude, we find that such effects will convert the sign of mixing-induced \CP violation $S_f$ into the positive one for the color-suppressed decays $\bar B_s\to K_S(\pi^0,\eta,\eta')$. Therefore, even the  measurements of the sign of $S_{\bar B_s\to K_S(\pi^0,\eta,\eta')}$ will be helpful to test if ``$a_2$" has a large magnitude and strong phase.

\item
Transverse polarization effects are sizable in penguin-dominated $\bar B_s\to VV$ as expected. However, the pQCD approach predicts that $f_L\sim f_T\sim {1\over 2}$ even for the color-suppressed tree-dominated decays $\bar B_s\to K^{*0}(\rho^0,\omega)$. This should be tested by experiment.


\end{enumerate}

\vskip 1.5in
\acknowledgments We are grateful to Cai-Dian L\"u and Amarjit Soni for discussions. One of us (HYC) wishes to thank the
hospitality of the Physics Department, Brookhaven National
Laboratory.
 This research was supported in part by the
National Science Council of R.O.C. under Grant Nos.
NSC97-2112-M-001-004-MY3 and NSC97-2112-M-033-002-MY3 and by the
National Center for Theoretical Science.



 \end{document}